\documentclass[fleqn,final,authoryear,3p,times,twocolumn]{elsarticle}

\usepackage{amsmath, amssymb, amsfonts}
\usepackage{color, multirow, multicol, array}
\usepackage{graphicx}
\usepackage{natbib}
\usepackage{charter}
\usepackage[english]{babel}

\pdfoutput=1        

\def\dst{\displaystyle}        
\def\const{{\rm const}}        
\def\cos{\mathop{\rm cos}\nolimits}
\def\sin{\mathop{\rm sin}\nolimits}

\def\max{{\rm max}}
\def\minmod{{\rm minmod}}
\def\sign{{\rm sign}}
\def\rot{\nabla \times}

\def\U{{\cal U}}
\def\F{{\cal F}}
\def\G{{\cal G}}

\def\B{{\bf B}}
\def\E{{\bf E}}

\def\dfrac#1#2{\frac{\displaystyle#1}{\displaystyle#2}}

\def\prodi#1#2{\frac{d #1}{d #2}}

\def\parti#1#2{\frac{\partial #1}{\partial #2}}
\def\partii#1#2{\frac{\partial^2 #1}{\partial {#2}^2}}

\renewcommand{\sectionmark}[1]{}
\renewcommand{\subsectionmark}[1]{}
\newcommand{\e}[1]{\times 10^{#1}}

\newcommand{\fig}[1]{Fig. \ref{#1}}

\journal{New Astronomy}    

\begin{document}

\begin{frontmatter}



\title{Numerical MHD Codes for Modeling Astrophysical Flows}


\author[1,2]{A. V. Koldoba}
\author[3]{G. V. Ustyugova}
\author[4]{P. S. Lii}
\author[5]{M. L. Comins}
\author[6]{S. Dyda}
\author[7]{M. M. Romanova}
\author[8,9]{R. V. E. Lovelace}

\address[1]{Moscow Institute of Physics and Technology, Dolgoprudnyy, Moscow region, 141700, Russia, email:
koldoba@rambler.ru}
\address[2]{Institute of Computer Aided Design RAS, $2^{nd}$ Brestskaya st., 19/18, Moscow, 123056, Russia}
\address[3]{Keldysh Institute of Applied Mathematics RAS,
Miusskaya sq., 4, Moscow, 125047, Russia, email: ustyugg@rambler.ru}
\address[4]{Department of Astronomy, Cornell University, Ithaca, NY 14853, email:psl46@cornell.edu}
\address[5]{Department of Astronomy, Cornell University, Ithaca, NY 14853, email:megan.comins@gmail.edu}
\address[6]{Department of Physics, Cornell University, Ithaca, NY 14853, email:sd449@cornell.edu}
\address[7]{Department of Astronomy, Cornell University, Ithaca, NY 14853, email:romanova@astro.cornell.edu,~ tel:
+1(607)255-6915, ~fax: +1(607)255-3433}
\address[8]{Department of Astronomy, Cornell University, Ithaca, NY 14853, email:lovelace@astro.cornell.edu}
\address[9]{Also Department of Applied and Eng. Physics}

\begin{abstract}
We describe a Godunov-type magnetohydrodynamic (MHD) code based on the Miyoshi \& Kusano (2005) solver which can be used to solve various astrophysical hydrodynamic and MHD problems. The energy equation is in the form of entropy conservation. The code has been implemented on several different coordinate systems: 2.5D axisymmetric cylindrical coordinates, 2D Cartesian coordinates, 2D plane polar coordinates, and fully 3D cylindrical coordinates.  Viscosity and diffusivity are implemented in the code to control the accretion rate in the disk and the rate of penetration of the disk matter through the magnetic field lines. The code has been utilized for the numerical investigations of a number of different astrophysical problems, several examples of which are shown.

\end{abstract}


\end{frontmatter}


\section{Introduction}
\label{sec:introduction}

A number of numerical magnetohydrodynamic (MHD) codes have been developed for modeling plasma flows in astrophysics. Some of the most well-known codes are \textit{ZEUS} \citep{ZEUS1992}, \textit{FLASH} \citep{FryxellEtAl2000}, \textit{PLUTO} \citep{MignoneEtAl2007}, and \textit{ATHENA} \citep{StoneEtAl2008,SkinnerOstriker2010}. A number of different numerical algorithms have also been developed for the numerical integration of the MHD equations including different approaches for the spatial and temporal approximations \citep{BW, Cockburn1989, Dai1994a, Dai1994b, RyuJones1995, BalsaraSpicer1999, Gurski2004, UstyugovEtAl2009}, and different algorithms for the approximate solution of the Riemann problem \citep{BW, Li2005, MiyoshiKusano2005,MiyoshiEtAl2010}.

In this work, we describe a code developed by our group for the numerical modeling of astrophysical MHD flows. The code has been developed for use in several coordinate systems: (1) 2.5D axisymmetric cylindrical coordinates ($r, z$); (2) 2D polar coordinates ($r, \phi$); (3) 3D cylindrical coordinates ($r, \phi, z$); and (4) a Cartesian ($x$, $y$) geometry which is used to conduct tests of the ideal and non-ideal MHD modules. The difference between our code and the above-mentioned codes lies in the specifics of the astrophysical problems that we solve.  Firstly, in the astrophysical regimes that we study, strong shocks (where the energy dissipation cannot be neglected) are not expected to occur.  This permits the use of the entropy conservation equation instead of the full energy equation.  The advantage of this approach is that the entropy conservation equation does not contain terms that differ significantly in magnitude. For example, in the energy equation, the largest terms are the gravitational energy and the kinetic energy of the azimuthal motion; these can be much larger than the internal energy and the energy of the poloidal motion near a gravitating body (for example, a planet or star). Additionally, in the vicinity of a magnetized star, the magnetic energy density can significantly exceed the matter pressure (and the energy-density of thermal energy), leading to significant errors when computing the energy conservation equation.

The different geometries have each been developed to solve a specific astrophysical problem and hence,  there are slight differences in the algorithms across the various coordinate systems. The first version of the code to be developed was the axisymmetric 2.5D version; its aim is to model the interaction of accreting magnetized stars with the surrounding accretion disk.  The 2D version in polar coordinates was developed later on to study the situation where planets orbit in a magnetized disk. Lastly, the 3D cylindrical version of the code was created to study intrinsically three dimensional problems such as bending waves in the disk or planets on inclined orbits. Each version of the code is based on the standard ideal MHD approach: the matter flow can be described with the one-fluid approximation.

In the 2D polar version of the code, the components of the velocity and magnetic field perpendicular to the plane of the flow are set to zero. Additionally, surface density is used instead of volume density.  The formal structure of the MHD equations is the same but we make additional suggestions about the disk and the definitions of the (surface) magnetic field so that the conservation equations retain the same conservative form. Our codes use a Riemann solver based on methods developed by \citet{MiyoshiKusano2005}, modified to include the equation for the entropy balance.

The 2.5D axisymmetric cylindrical version of the code was developed to investigate astrophysical processes like disk accretion, magnetospheric accretion onto a magnetized star and outflows launched from the disk and star. Such MHD flows may be responsible for many the observed properties of young stars. In all of these processes, both the magnetic field of the star as well as magnetic fields induced in the magnetosphere of the star and in the accretion disk play an important role. It is often suggested that the angular momentum transport in the accretion disk is due to the magnetic turbulence associated with the magnetorotational instability (MRI) (e.g., \citealt{BalbusHawley1991}). However, modelling this instability requires high grid resolution in the disk, which is computationally expensive. In many cases, it is more practical to allow for steady accretion by incorporating viscosity terms into the code that mimic the angular momentum transport, but do not require high grid resolution. For this reason, the 2.5D version incorporates an $\alpha$-type viscosity \citep{ShakuraSunyaev1973} which can be switched on or off depending on the phenomena being studied.

It is less clear how the matter in the disk interacts with magnetic field--that is, which physical processes provide efficient magnetic diffusivity in the disk. Reconnection of the magnetic field lines can provide an efficient diffusion mechanism, but the process of reconnection itself also requires some magnetic diffusivity. Since the exact mechanism is poorly understood, the magnetic diffusivity is often parametrized in the non-ideal MHD terms. This is implemented in the 2.5D version of the code as a diffusivity module, which can also be switched on or off, with the diffusivity coefficient represented in a way analogous to the $\alpha-$viscosity.

In the 2.5D axisymmetric cylindrical version of the code, we split the magnetic field into a fixed component associated with the stellar magnetic field (for example a dipole field) and a field induced by currents in the magnetosphere and disk \citep{Tanaka1994}.  In contrast, the magnetic field in the 2D polar version is not split into a stellar and current-induced component as the code was initially developed to study the interaction of a planet with a disk threaded by a magnetic field. In order to accurately compute the strength of the stellar gravity on the planet, the grid is allowed to co-rotate with the planet. This approach was suggested by \citep{Kley1998}, albeit for a different situation. 

The 3D cylindrical version of the code combines the two previous versions, enabling investigations of phenomena such as planet migration in a magnetized accretion disk, accretion onto a magnetized star as well as a variety of other problems. However, there is no diffusivity module implemented in the 3D version because the 3D instabilities which cause the magnetic field to diffuse are fully modelled.

In this work, we describe our Godunov-type codes which have been implemented on several different geometries and are designed for solving non-relativistic astrophysical MHD problems. Sec. \ref{sec:Godunov} describes different Godunov approaches and the approach used in our codes. In Sec. \ref{sec:numerical algorithms}, we describe the different coordinate systems in detail.  In Sec. \ref{sec:tests}, we describe tests of the code. Sec. \ref{sec:astrophysical examples} shows applications of these codes for different astrophysical problems and we conclude with a summary in Sec. \ref{sec:conclusions}.

\section{The governing equations of MHD in cylindrical coordinates}

Here we present the full set of ideal 3D MHD equations. Viscosity and diffusivity are implemented for specific geometries and are described separately in Sec. \ref{subsec:viscosity and diffusivity}. The continuity equation in conservative form is:
\begin{equation} \label{eq:cons_eqn}
\parti{\rho}{t} + \frac{1}{r} \parti{(r \rho v_r)}{r} + \frac{1}{r}
\parti{(\rho v_\phi)}{\phi} + \parti{(\rho v_z)}{z} = 0 ~.
\end{equation}
Here $\mathbf{r} = (r, \phi, z)$ are the 3D cylindrical coordinates, $\rho$ is the density, $\mathbf{v} = (v_r, v_\phi, v_z)$ is the velocity, and $t$ is the time. The entropy conservation equation is:
\begin{equation}
\parti{(\rho s)}{t} + \frac{1}{r} \parti{(r \rho s v_r)}{r} + \frac{1}{r} \parti{(\rho s v_\phi)}{\phi} + \parti{(\rho
s v_z)}{z} = 0 ~,
\end{equation}
where $s \equiv p/\rho^\gamma$ is the entropy per unit mass and $\gamma$ is the adiabatic index. The momentum equations in the $r$, $\phi$ and $z$ directions are
\begin{align} \label{eq:momeqn_r}
&\parti{(\rho v_r)}{t} + \parti{}{r} \left( \rho v_r^2 + Q - \frac{B_r^2}{4\pi} \right) + \nonumber \\
&\frac{1}{r} 
\parti{}{\phi} \left( \rho v_r v_\phi - \frac{B_r B_\phi}{4\pi}
\right) + \parti{}{z} \left( \rho v_r v_z - \frac{B_r
B_z}{4\pi} \right) = \nonumber \\
&\frac{1}{r} \left( \rho (v_\phi^2-v_r^2) - \frac{B_\phi^2-B_r^2}{4\pi} \right) + \rho g_r ~,
\end{align}
\begin{align} \label{eq:momeqn_phi}
&\parti{(\rho v_\phi)}{t} + \frac{1}{r^2} \parti{}{r} r^2 \left( \rho v_r v_\phi - \frac{B_r B_\phi}{4\pi} \right) + \nonumber \\
&\frac{1}{r} \parti{}{\phi} \left( \rho v_\phi^2 + Q - \frac{B_\phi^2}{4\pi} \right) +
\parti{}{z} \left( \rho v_\phi v_z -
\frac{B_\phi B_z}{4\pi} \right) = 0 ~,
\end{align}
\begin{align} \label{eq:momeqn_z}
\parti{(\rho v_z)}{t} + & \frac{1}{r} \parti{}{r} r \left( \rho v_r v_z - \frac{B_r B_z}{4\pi} \right) + \nonumber \\
& \frac{1}{r} \parti{}{\phi} \left( \rho v_\phi v_z - \frac{B_\phi B_z}{4\pi} \right) + \nonumber \\
& \parti{}{z} \left( \rho v_z^2 + Q - \frac{B_z^2}{4\pi} \right) = \rho g_z ~,
\end{align}
where $\mathbf{B} = (B_r, B_\phi, B_z)$ is the magnetic field vector, $Q = p + B^2/8\pi$ is the total pressure, and $\mathbf{g} = - \nabla \Phi$ is the net external force from the central star and planet. Lastly, the induction equations are:
\begin{align} \label{eq:full-3d}
\parti{B_r}{t} + & \frac{1}{r} \parti{}{\phi} \left( v_\phi B_r - v_r B_\phi \right) + \parti{}{z} \left( v_z B_r - v_r B_z \right) = 0 \\
\parti{B_\phi}{t} + & \parti{}{r} \left( v_r B_\phi - v_\phi B_r \right) + \parti{}{z} \left( v_z B_\phi - v_\phi B_z \right) = 0 \\
\parti{B_z}{t} + & \frac{1}{r} \parti{}{r} r \left( v_r B_z - v_z B_r \right) + \frac{1}{r} \parti{}{\phi} \left( v_\phi B_z - v_z B_\phi \right) = 0 
\end{align}
This system of equations describes the evolution of a magnetized plasma in cylindrical coordinates.

\section{Godunov type schemes}
\label{sec:Godunov}
In order to numerically solve the ideal MHD equations (Eqs. \ref{eq:cons_eqn}-\ref{eq:full-3d}), we use the Godunov method. In Godunov-type schemes, the central problem is to develop an exact or approximate algorithm to solve the Riemann problem. The construction of an exact solution to the Riemann problem is possible, but it can be numerically complex and is often computationally expensive. The procedure of exact solutions of the Riemann problem has been realized by \citet{RyuJones1995} and these exact solutions are used for the testing of our approximate Riemann solver.

Consider the one-dimensional hyperbolic system of equations in conservative form:
\begin{equation} \label{1-1}
\parti{\U}{t} + \parti{\F(\U)}{x} = 0.
\end{equation}
Here, $\U = \{ U_1,..., U_n \}$ is the vector of conservative variables and ${\F(\U)} = \{ F_1,..., F_n \}$ is the flux function which maps each conservative variable to its corresponding flux. Eq. \ref{1-1} can be rewritten in the integral form of the conservation laws
\begin{equation} \label{1-1a}
\int_{\Delta\xi} \U_{ex}(\xi) d\xi = \int_{\Delta\xi} \U|_{t=0} d\xi + \F_L - \F_R,
\end{equation}
where $\U_{ex}(\xi;\U_L,\U_R)$ represents the \textit{exact solution} of the Riemann problem between the left and right states $U_L$ and $U_R$; $\xi = x/t$ is a self-similar distance-time variable; $\Delta\xi$ is the interval along the variable $\xi$ at which all of the waves are localized; and $\F_L=\F(\U_L)$ and $\F_R=\F(\U_R)$ are the fluxes for the left and right states, respectively.

The Riemann problem can be approximated by a number of discontinuities, each of which satisfies the Rankine-Hugoniot jump conditions:
\begin{equation} \label{1-1d}
\dot x\ (\U_2 - \U_1) = \F_2 - \F_1.
\end{equation}
Here, $\dot x$ is velocity of the discontinuity; $\U_1, \U_2$ are the conservative variables on either side of the discontinuity; and $\F_1,\F_2$ are the fluxes of the conservative variables across the discontinuity. In the general case, $\F_1 \ne \F(\U_1)$ and $\F_2 \ne \F(\U_2)$, excluding $\F_L = \F(\U_L), \F_R = \F(\U_R)$. In this case, the condition for self-consistency of the conservation laws is evidently satisfied, while Eq. \ref{1-1b} for the calculation of fluxes gives $\F=\F^*$ , where $\F^*$ is the flux in the interval which includes the point $x=0$. If the point $x = 0$ is located at the boundary between intervals (that is, at the discontinuity where $\dot x = 0$), then the flux can be calculated for any two adjacent states. These fluxes are the same as the ones present in the Rankine-Hugoniot jump conditions.

\begin{figure}
\centering
\includegraphics[width=8.0cm]{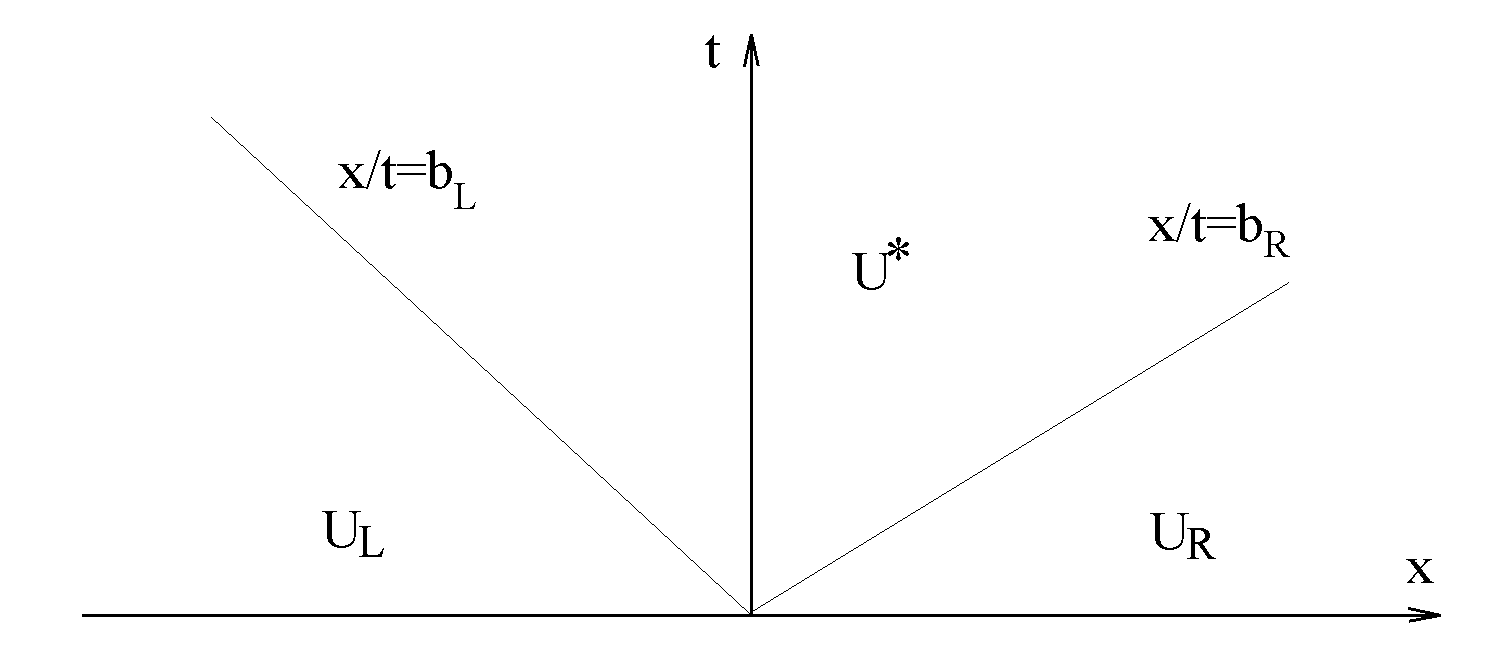}
\caption{The wave propagation diagram for the HLL method in the $(x,t)$-plane.} \label{hll}
\end{figure}

The approximate solution of the Riemann problem is represented by the vector $\U(\xi)$ which approximates the same conservation laws in integral form. The flux between the left and right states is approximated by
\begin{align}
\label{1-1b}  FLUX(U_L,\U_R) & = \F_L  + \int_0^\infty ( \U(\xi) - \U_L) d \xi \nonumber \\ 
&= \F_R  - \int_0^\infty ( \U(\xi) - \U_R) d \xi ~.
\end{align}
One of the simplest approximate Riemann solvers was proposed by \citet{HartenEtAl1983} and has two discontinuities separating three homogeneous states (as shown in \fig{hll}).  One discontinuity propagates to the left with velocity $b_L$, while the other propagates to the right with velocity $b_R$. The approximate solution is assumed to be homogeneous between these discontinuities: $\U(\xi) = \U^*$.

\begin{figure}
\centering
\includegraphics[width=7.8cm]{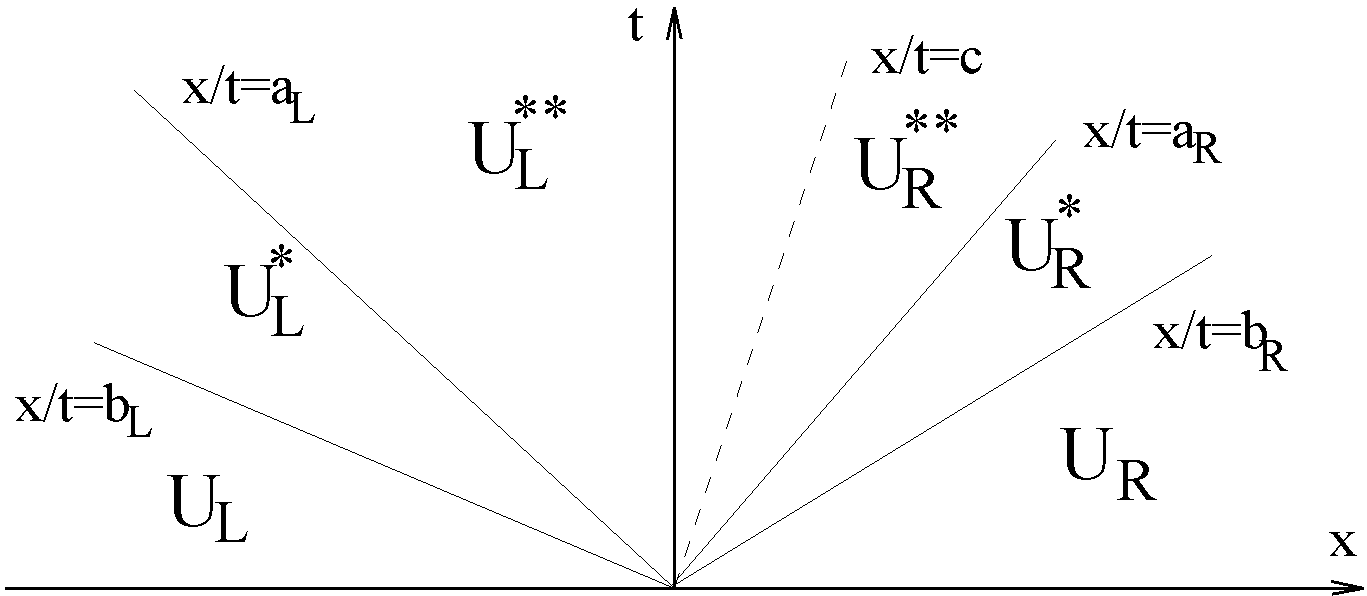}
\caption{The wave propagation diagram for the HLLD method in
the $(x,t)$-plane.} \label{hlld}
\end{figure}
More advanced approximate Riemann solvers include two or more intermediate states \citep{Gurski2004, Li2005, MiyoshiKusano2005}.

\subsection{The Miyoshi \& Kusano HLLD Solver} \label{subsec:MiyoshiKusanoSolver}
Here, we describe the construction of our Godunov method for the equations of ideal MHD. As noted earlier, we use the entropy conservation equation instead of the energy equation. The conservative variables and fluxes in Eq. \ref{1-1} take on the form:
\begin{align*}
\U = & \{\rho,~ \rho s,~ \rho v_x,~ \rho v_y,~ \rho v_z,~ B_y,~
B_z \}, \\
\F = & \{\rho v_x,~ \rho s v_x,~ \rho v_x^2 + p + \frac{B_y^2 + B_z^2}{4 \pi}, \\
& \rho v_y v_x - \dfrac{B_{x0} B_y}{4\pi},~ \rho v_z v_x - \dfrac{B_{x0} B_z}{4 \pi}, \\
& v_x B_y - v_y B_{x0}, v_x B_z - v_z B_{x0} \} 
\end{align*}

\citet{Li2005} proposed a modification of the HLLC approximate Riemann solver for the equations of MHD in which the transverse $y, z$-components of the velocity and magnetic field are assumed to be the same in both intermediate states.  For this reason, in the limit $B_{x0} \to 0$, the HLLC solver does not convert to the HLLC algorithm with a purely transverse magnetic field; this is because arbitrary jumps of the $y,z$-components of velocity and magnetic field are still allowed at the contact discontinuity (as they are converted to the tangential component).

The choice of entropy conservation inhibits the modeling of MHD flows with strong shocks at which entropy production occurs. Our  approximate solution has four intermediate states---the wave propagation in the HLLD algorithm is shown in \fig{hlld}. The initial discontinuity between the states $\U_L$ and $\U_R$ decays and the discontinuities appear, propagating away with velocities $b_L, b_R$ (the fast magnetosonic waves), $a_L, a_R$ (the Alfv\'en waves) and $c$ (the contact discontinuity). These jumps separate the initial $\U_L, \U_R$ and intermediate $\U_L^*, \U_L^{**}, \U_R^*, \U_R^{**}$ states (see \fig{hlld}).  The calculation of the conservative variables and fluxes in the intermediate states is performed using the following scheme. The normal velocities across the Alfv\'en and contact discontinuities are assumed to be continuous
\begin{equation*}
v_{xL}^* = v_{xL}^{**} = v_{xR}^* = v_{xR}^{**} = c,~~~~
\end{equation*}
which implies
\begin{equation*}
\rho_L^* = \rho_L^{**},~~~~\rho_R^* = \rho_R^{**},~~~~s_L^* = s_L^{**},~~~~s_R^* = s_R^{**},
\end{equation*}
from the Rankine-Hugoniot jump conditions.

We write the Rankine-Hugoniot conditions for fast magnetosonic waves for all equations except the conservation equation for the $x-$component of the momentum. For the densities in states $\U_L^*$ and $\U_R^*$, we obtain the relations
\begin{align} \label{1-2} 
\rho_L^* &= \rho_L \frac{b_L - v_{xL}}{b_L - v_{xL}^*} = \rho_L \frac{b_L - v_{xL}}{b_L - c}, \nonumber \\
\rho_R^* &= \rho_R \frac{b_R - v_{xR}}{b_R - v_{xR}^*} = \rho_R \frac{b_R - v_{xR}}{b_R - c}.
\end{align}
From Eq. \ref{1-2} and the $x$-component of the momentum equation we obtain
\begin{equation} \label{1-3}
c = \frac{\chi}{\rho_L (v_{xL} - b_L) +\rho_R (b_R - v_{xR})},
\end{equation}
where
\begin{align*}
\chi \equiv &\left[ \left( \rho v_x^2 + p + \frac{B^2}{8\pi} - \frac{B_{x0}^2}{4\pi}\right)_L - b_L (\rho v_x)_L \right] - \\
& \left[ \left( \rho v_x^2 + p + \frac{B^2}{8\pi} - \frac{B_{x0}^2}{4\pi} \right)_R - b_R (\rho v_x)_R \right] ~.
\end{align*}

Combining Eq. \ref{1-2} and the Rankine-Hugoniot jump conditions for the entropy, we find
$$
s_L^* = s_L^{**} = s_L,~~~~s_R^* = s_R^{**} = s_R ~.
$$

To calculate the transverse components of the velocity and magnetic field in the intermediate states, we use the corresponding components of the momentum and induction equations. In the intermediate states, $\U_L^*$ and $\U_R^*$, these values are determined from the jump conditions across the fast magnetosonic waves:
\begin{align} \label{1-4}
v_{yL}^* &= v_{yL} + \frac{B_{x0} B_{yL}}{4\pi} \dfrac{v_{xL}-c}{\rho_L^*(b_L-c)^2-\dfrac{B_{x0}^2}{4\pi} }, \nonumber \\
B_{yL}^* &= B_{yL} \dfrac{\rho_L (b_L-v_{xL})^2 - \frac{B_{x0}^2}{4\pi} } {\rho_L^*(b_L-c)^2 - \dfrac{B_{x0}^2}{4\pi}}.
\end{align}
Analogously, we have
\begin{align} \label{1-5} 
v_{zL}^* &= v_{zL} + \frac{B_{x0} B_{zL}}{4\pi} \dfrac{v_{xL}-c}{\rho_L^*(b_L-c)^2-\dfrac{B_{x0}^2}{4\pi} }, \nonumber\\
B_{zL}^* &= B_{zL} \dfrac{\rho_L (b_L-v_{xL})^2 - \frac{B_{x0}^2}{4\pi}} {\rho_L^*(b_L-c)^2-\dfrac{B_{x0}^2}{4\pi}}~.
\end{align}
The values of $v_{yR}^*, v_{zR}^*, B_{yR}^*, B_{zR}^*$ are calculated using the same formulae, substituting the index $L \to R$.

The Rankine-Hugoniot jump conditions at the Alfv\'en waves (with $\rho_L^{**} = \rho_L^*,~~\rho_R^{**} = \rho_R^*$) can be solved, but only if
$$
a_L =c \pm \dfrac{B_{x0}}{\sqrt{4\pi \rho_L^*}},~~~~a_R =c \pm \dfrac{B_{x0}}{\sqrt{4\pi \rho_R^*}} .
$$
In the first case, we adopt $a_L = c - \dfrac{|B_{x0}|}{\sqrt{4\pi \rho_L^*}}$, which corresponds to an Alfv\'en wave propagating to the left along the state $\U_L^*$. In the second case, we take $a_R = c + \dfrac{|B_{x0}|}{\sqrt{4\pi \rho_R^*}}$, which corresponds to an Alfv\'en wave propagating to the right along the state $\U_R^*$.

To compute the transverse components of the velocity and magnetic field in the intermediate states $\U_L^{**}, \U_R^{**}$, we use the fact that at the contact discontinuity
\begin{align*}
v_{yL}^{**} &= v_{yR}^{**} = v_y^{**},~~~~  v_{zL}^{**} = v_{zR}^{**} = v_z^{**}, \\
B_{yL}^{**} &= B_{yR}^{**} = B_y^{**},~~ B_{zL}^{**} = B_{zR}^{**} = B_z^{**},
\end{align*}
for $B_{x0} \ne 0$.

Plugging these into the integral conservation laws, we find that the transverse components of the velocity and magnetic field are
\begin{align*}
&v_y^{**} = \dfrac{v_{yL}^* \sqrt{\rho_L^*} + v_{yR}^* \sqrt{\rho_R^*} + \sigma (B_{yR}^* - B_{yL}^*)/\sqrt{4\pi}} {\sqrt{\rho_L^*} + \sqrt{\rho_R^*} }, \\
&v_z^{**} = \dfrac{v_{zL}^* \sqrt{\rho_L^*} + v_{zR}^* \sqrt{\rho_R^*} + \sigma (B_{zR}^* - B_{zL}^*)/\sqrt{4\pi}} {\sqrt{\rho_L^*} + \sqrt{\rho_R^*} }, \\
&B_y^{**} = \dfrac{B_{yL}^* \sqrt{\rho_R^*} + B_{yR}^* \sqrt{\rho_L^*} + \sigma \sqrt{4\pi \rho_L^* \rho_R^*} (v_{yR}^* - v_{yL}^*)} {\sqrt{\rho_L^*} + \sqrt{\rho_R^*} }, \\
&B_z^{**} = \dfrac{B_{zL}^* \sqrt{\rho_R^*} + B_{zR}^* \sqrt{\rho_L^*} + \sigma \sqrt{4\pi \rho_L^* \rho_R^*} (v_{zR}^* - v_{zL}^*)} {\sqrt{\rho_L^*} + \sqrt{\rho_R^*} } ~,
\end{align*}
where $\sigma = \sign(B_{x0})$ \citep{MiyoshiKusano2005}.

Note that the transverse components of the magnetic field $B_t$ are not necessarily equal ($|B_{tL}^*| \ne |B_t^{**}| \ne |B_{tR}^*|$) due to the jump conditions for the Alfv\'en waves in the complete system of MHD equations. Here, we remove the condition $\F^* = \F(\U^*)$ (in this case, for the $x-$component of the momentum equation) and hence lose the continuity of the magnetic pressure for the Alfv\'en waves.

The fluxes in the intermediate states are calculated from the jump conditions
\begin{align}
\F_L^*   &= \F_L + b_L ( \U_L^* - \U_L ), \label{1-5a}\\
\F_L^{**} &= \F_L^* + a_L ( \U_L^{**} - \U_L^* ) \nonumber\\ 
& = \F_L + a_L ( \U_L^{**} - \U_L^* ) + b_L ( \U_L^* - \U_L ), \label{1-5b}\\
\F_R^* & = \F_R + b_R ( \U_R^* - \U_R ), \label{1-5c} \\
\F_R^{**} & = \F_R^* + a_R ( \U_R^{**} - \U_R^* ) \nonumber\\ 
& = \F_R + a_R ( \U_R^{**} - \U_R^* ) + b_R ( \U_R^* - \U_R ). \label{1-5d}
\end{align}
One does not need to calculate all of the fluxes. Rather, we determine which of the states appears at the point $x = 0$ (in other words, which interval $(-\infty,b_L),...,(b_R,\infty)$ contains the point $x = 0$) and then calculate the corresponding flux using one of the formulae (\ref{1-5a})-(\ref{1-5d}).

In the limit $B_{x0} \to 0$, the relationships in Eq. \ref{1-4} give the result
$$
v_{yL}^* = v_{yL},~~~~ B_{yL}^* = B_{yL} \dfrac{\rho_L (b_L-v_{xL})^2}{\rho_L^*(b_L-c)^2} = B_{yL} \dfrac{\rho_L^*}{\rho_L}.
$$
Similar relationships can be derived for the $z$-component of velocity and magnetic field and for the fast magnetosonic wave which propagates to the right along $\U_R$. Additionally since $a_L = a_R = c$ when $B_{x0} = 0$, the intermediate states $\U_L^{**}, \U_R^{**}$ disappear.

In the above derivations it is assumed that the inequalities
$$
b_L \le a_L \le c \le a_R \le b_R
$$
are satisfied. The case $a_L = c = a_R$, corresponding to $B_{x0} = 0$, has been considered above. The cases where $b_L = a_L$ or $a_R = b_R$ correspond to the switch-on waves propagating to the left or to the right, assuming $B_{t,L} = 0$ or $B_{t,R}=0$ at $B_{x0} \ne 0$, and are not considered. The wave velocities $b_L$ and $b_R$ are estimated as
$$
b_L \le (v_x - c_F)_L,~~~~b_R \ge (v_x + c_F)_R,
$$
where $c_{F,L}, c_{F,R}$ are the velocities of the fast magnetosonic waves, propagating along the states $\U_L, \U_R$ respectively.

\subsection{Correction of wave velocities} \label{subsec:correction of waves}
Equations \ref{1-4} and \ref{1-5} show the values of the tangential velocities and magnetic fields in the intermediate states $\U_L^*, \U_R^*$.
The numerators on the right-hand sides of these formulae are non-negative, while the denominators have the form
\begin{equation*}
\Delta =  \rho_R^* (b_R - c)^2 - \dfrac{B_{x0}^2}{4\pi} = \rho_R
(b_R - v_{xR})(b_R - c)- \dfrac{B_{x0}^2}{4\pi}.
\end{equation*}
The formula for the left state is analogous.

The physical sense of the last value is such that it should be positive because the Alfv\'en wave should always propagate slower than the fast MHD discontinuity. The case when this denominator is zero corresponds to the switch-on wave, in which a finite tangential field beyond the discontinuity appears from the zero tangential field in front of the discontinuity. In the approximate Riemann solver it is reasonable to retain this property (the positive sign of the denominator) in order to avoid a non-physical change of the sign of the tangential field and the appearance of a singularity as $\Delta \to 0$. For that, we require that the inequality $b_R \ge c + \dfrac{k|B_{x0}|}{\sqrt{4\pi \rho_R^*}}$ is satisfied, at some $k>1$. In that case, we have
$$
b_R - c - \dfrac{k|B_{x0}|}{\sqrt{4\pi \rho_R^*}} =
b_R - c - \dfrac{k|B_{x0}| \sqrt{b_R-c}}{\sqrt{4\pi \rho_R (b_R-v_{xR})}}
\ge 0.
$$
This can be rewritten as
\begin{equation}
\label{1-6}
\rho_R (b_R-c)( b_R-v_{xR}) \ge \dfrac{k^2 B_{x0}^2}{4\pi} ~.
\end{equation}
Comparing this relationship with that for $\Delta$, we conclude that if the condition in Eq. \ref{1-6} is satisfied then $\Delta > 0$. Analogously:
\begin{equation}
\label{1-7} \rho_L (b_L-c)( b_L-v_{xL}) \ge \dfrac{k^2 B_{x0}^2}{4\pi} ~.
\end{equation}
If the conditions in Eq. \ref{1-6} and/or Eq. \ref{1-7} are not satisfied (after computing $c$ using Eq. \ref{1-3}) then we perform a correction of $b_R$ and/or $b_L$, choosing the fast magnetosonic velocity to be the largest and/or the smallest of the roots of Eqs. \ref{1-6} and/or \ref{1-7}. Here, $b_R$ can only increase while $b_L$ can only decrease. We then recalculate the value $c$, check the conditions in Eqs. \ref{1-6} and \ref{1-7}, and if necessary, once again apply corrections to $b_L$ and $b_R$.

\subsection{Positivity} \label{subsec:positivity}
We use the equation for entropy conservation instead of the equation of energy conservation. In this case, the correction of the velocities of the fast MHD discontinuities and the positivity of the approximate solution of the Riemann solver (in the sense that the density and entropy are positive in all intermediate states) is established trivially.

\subsection{Integration in time} \label{subsec:integration in time}
To increase the accuracy of the numerical algorithm, we perform a two-stage Runge-Kutta integration of the equations in time.  In the first stage, we calculate the values for the intermediary timestep, $\U_i^{n+1/2}$
$$
\U_i^{n+1/2} = \U_i^n - \dfrac{\Delta t}{2 \Delta x} (\F_{i+1/2}^n - \F_{i-1/2}^n).
$$ 
Here, the fluxes $\F_{i+1/2}^n$ are calculated using the entropy-based HLLD algorithm described above. Formally, we can write the flux calculation procedure in the form $\F_{i+1/2}^n = FLUX(\U_i^n, \U_{i+1}^n)$ where we take the left and right states in the HLLD solver to be $\U_L = \U_i$ and $\U_R = \U_{i+1}$.

In the second stage, the values for the full timestep, $t^{n+1} = t^n + \Delta t$, are calculated as
$$
\U_i^{n+1} = \U_i^n - \dfrac{\Delta t}{\Delta x} (\F_{i+1/2}^{n+1/2} - \F_{i-1/2}^{n+1/2}).
$$
The timestep $\Delta t$ is set by the Courant-Friedrichs-Levy (CFL) condition, which places an upper limit on the timestep by considering the wave crossing speeds in all cells on the grid. For our grid, the timesteps are typically limited by the strong mangetic fields in the cells near the star. For the calculation of the fluxes $\F_{i+1/2}^{n+1/2}$ we use values from the intermediary timestep $\U_i^{n+1/2}$, reconstructed on the edges of the calculated cells:
$$
\F_{i+1/2}^{n+1/2} = FLUX(\U_{R,i}^{n+1/2}, \U_{L,i+1}^{n+1/2}).
$$
The values $\U_L, \U_R$ are calculated at each grid cell boundary using one of the previously described approximate Riemann solvers. In order to increase the accuracy of our scheme, a slope limiter correction is applied to the primitive variables at the edges of each cell. For the sake of brevity, we define $d_{i-1/2} \equiv u_i - u_{i-1}$ and $d_{i+1/2} \equiv u_{i+1} - u_i$.
\begin{align}
&u_{L,i} = u_i - \frac{1}{2} \minmod(d_{i-1/2}, d_{i+1/2}),\nonumber\\
&u_{R,i} = u_i + \frac{1}{2} \minmod(d_{i-1/2}, d_{i+1/2}) .
\nonumber\\
\end{align}
$$
u_{L,i} = u_i - \left \{
\begin{array}{ll}
0 & if~ d_{i+1/2}d_{i-1/2} \le 0, \\[0.3cm]
\dst{\frac{\epsilon+2}{6 \epsilon} d_{i+1/2} }  &
if~|d_{i+1/2}|  < \epsilon |d_{i-1/2}|, \\[0.3cm]
\dst{\frac{d_{i-1/2}}{3} + \frac{d_{i+1/2}}{6} }  &
if~\epsilon |d_{i-1/2}| \le |d_{i+1/2}|  \le 4 |d_{i-1/2}|, \\[0.3cm]
d_{i-1/2} & if~~~~|d_{i+1/2}| \ge 4 |d_{i-1/2}|~.
\end{array}
\right.
$$
$$
u_{R,i} = u_i + \left \{
\begin{array}{ll}
0 & if ~ d_{i+1/2}d_{i-1/2} \le 0, \\[0.3cm]
\dst{\frac{\epsilon+2}{6 \epsilon} d_{i-1/2} }  &
if ~ |d_{i-1/2}| < \epsilon |d_{i+1/2}|, \\[0.3cm]
\dst{\frac{d_{i+1/2}}{3} + \frac{d_{i-1/2}}{6} } &
if ~ \epsilon |d_{i+1/2}| \le |d_{i-1/2}| \le 4 |d_{i+1/2}|, \\[0.3cm]
d_{i+1/2} & if |d_{i-1/2}| \ge 4 |d_{i+1/2}| .
\end{array}
\right.
$$
Here, the top indicies are dropped and we take $\epsilon = 0.5$. In our algorithm, we reconstruct the primitive variables $p, s, v_x, v_y, v_z, B_x, B_y, B_z$.

\section{Numerical algorithms in cases of different geometries}
\label{sec:numerical algorithms}

\subsection{Cartesian geometry}
\label{subsec:geom-cartesian-coord}

In a 2D Cartesian $(x, y)$ geometry, the MHD equations take the form
$$
\parti{\U}{t} + \parti{\F}{x} + \parti{\G}{y} = 0~.
$$
Unlike the 1D case, the vector of the conservative variables $\U$ has a total of seven components. The sixth and
seventh components are $B_x$ and $B_y$, which must satisfy the divergence free condition
\begin{equation} \label{2-1}
\parti{B_x}{x} + \parti{B_y}{y} = 0.
\end{equation}
The numerical algorithm for 2D is analogous to the 1D case. The integration in time is performed using the two-step Runge-Kutta method
\begin{align*}
\U_{i,j}^{n+1/2} = \U_{i,j}^n &- \dfrac{\Delta t}{2 \Delta x} (\F_{i+1/2,j}^n - \F_{i-1/2,j}^n) ~ \\ 
&- \dfrac{\Delta t}{2 \Delta y} (\G_{i,j+1/2}^n - \G_{i,j-1/2}^n), \\
\U_{i,j}^{n+1}  = \U_{i,j}^n &- \dfrac{\Delta t}{\Delta x} (\F_{i+1/2,j}^{n+1/2} - \F_{i-1/2,j}^{n+1/2}) \\
&- \dfrac{\Delta t}{\Delta y} (\G_{i,j+1/2}^{n+1/2} - \G_{i,j-1/2}^{n+1/2}).
\end{align*}
At the same time, $\F_{i+1/2,j}^n = FLUX(\U_{i,j}^n, \U_{i+1,j}^n),~~\G_{i,j+1/2}^n = FLUX(\U_{i,j}^n, \U_{i,j+1}^n)$.  An analogous algorithm for the calculation of fluxes is also used during the second stage of the time-advancement computation.

To ensure that the magnetic field is divergence-free in 2D, we use the method proposed by \citet{BalsaraSpicer1999}. In application to problems in Cartesian coordinates, this approach is the following: if Eq. \ref{2-1} is satisfied, then one can represent the $x$ and $y-$components of the magnetic field in the form
$$B_x = \parti{A_z}{y},~~~~ B_y = - \parti{A_z}{x}.$$
At the same time, $\dst \parti{A_z}{t} = - E_z$. When the equations are discretized, the conservative variables $\U$ are determined in the cells while the $x$ and $y$ fluxes, $\F$ and $\G$, are computed on the sides of the cells. The corresponding components of the flux function vectors $\F$ and $\G$ are
$$
F_6 = 0,~~F_7 = - E_z,~~G_6 = E_z,~~G_7 = 0,
$$
where $E_z = - (v_x B_y - v_y B_x)$ is the $z$ component of the electric field (multiplied by the speed of light).  Thus, the $z$ component of the electric field is calculated twice using two different methods: the first as $-F_{7,i+1/2,j}$, and the second as $G_{6,i,j+1/2}$. If $A_z$ is determined on the nodes of the grid $(i+1/2,j+1/2)$, the components of the magnetic field can be calculated on the sides of the cells as
$$
\bar{B}_{x,i+1/2,j} = \dfrac{A_{z,i+1,j} - A_{z,i,j}}{\Delta
y},~~~~
\bar{B}_{y,i,j+1/2} = - \dfrac{A_{z,i,j+1} -
A_{z,i,j}}{\Delta x}.
$$
At the same time, the divergence-free condition will be satisfied if we require
$$
\dfrac{\bar{B}_{x,i+1/2,j} - \bar{B}_{x,i-1/2,j}}{\Delta x} +
\dfrac{\bar{B}_{y,i,j+1/2} - \bar{B}_{y,i,j-1/2}}{\Delta y} = 0.
$$
The magnetic field in the cells are then computed from the calculated values on the cell boundaries
\begin{align} \label{2-1a} 
B_{x,i,j} = & \frac{1}{2}(\bar{B}_{x,i-1/2,j} + \bar{B}_{x,i+1/2,j} ), \nonumber\\
B_{y,i,j} = & \frac{1}{2}(\bar{B}_{y,i,j-1/2} + \bar{B}_{y,i,j+1/2} ) ~.
\end{align}

To calculate $A_{z,i+1/2,j+1/2}$ at the $n+1^{\rm th}$ timestep, we determine the $z-$component of the electric field at each node. The simple variant proposed by \citet{BalsaraSpicer1999} consists of averaging the values $-F_7$ and $G_6$ over the cell sides surrounding the node at $(i+1/2,j+1/2)$. For a homogeneous grid, this is
\begin{align} \label{2-2} 
&E^{BS}_{z,i+1/2,j+1/2} = \nonumber\\
&\frac{1}{4} (E_{z,i,j+1/2} + E_{z,i+1,j+1/2} + E_{z,i+1/2,j} + E_{z,i+1/2,j+1}) = \nonumber\\
&\frac{1}{4} (G_{6,i,j+1/2} + G_{6,i+1,j+1/2} - F_{7,i+1/2,j} - F_{7,i+1/2,j+1}) ~.
\end{align}

In \cite{GardinerStone2005} it was noted that this approach is inconsistent with the numerical integration algorithm in the case of 1D problems (where the solution does not depend either on $x$ or on $y$), and that the proposed procedure does not guarantee consistency.  \cite{GardinerStone2005} propose a modified form (for homogenous grids)
\begin{align}
\label{2-3} &~E^{GS}_{z,i+1/2,j+1/2} = 2 E^{BS}_{z,i+1/2,j+1/2}  \nonumber \\
&- \frac{1}{4} \left( E_{z,i,j} + E_{z,i+1,j} + E_{z,i+1,j+1} +
E_{z,i,j+1} \right). \end{align}
To calculate $E_z$, we use either Eqs. \ref{2-2} or \ref{2-3}. Eq. \ref{2-3} has applications in damping out small-scale oscillations of the magnetic field.

Thus, at each stage of numerical integration of the MHD equations, the magnetic field in the $(x,y)$ plane is calculated along the following algorithm:
\begin{enumerate}
\item Use an approximate Riemann solver to find the $z-$components of the electric field at the sides of cells $-F_{7,i+1/2,j}$, $G_{6,i,j+1/2}$.
\item Use Eq. \ref{2-2} or \ref{2-3} to calculate the electric field in the nodes of the grid $E_{z,i+1/2,j+1/2}$.
\item Calculate the $z-$component of the magnetic field potential in grid nodes, $A_{z,i+1/2,j+1/2}$.
\item Calculate  $\bar{B}_{x,i+1/2,j}$ and $\bar{B}_{y,i,j+1/2}$.
\item Using Eq. \ref{2-1a}, calculate $B_{x,i,j}$ and $B_{y,i,j}$.
\end{enumerate}

\subsection{Axisymmetric 2.5D cylindrical geometry} \label{subsec:geom-cyl-coord} 
To solve the 2.5D MHD equations (with magnetic diffusivity and viscosity) in the disk, we split the different physical processes into several different modules. These are: (1) the ``ideal MHD" module in which we calculate the dynamics of the plasma and magnetic field with dissipative processes switched off; the ``diffusivity'' modules (2) and (3) which calculate the diffusion of the poloidal and azimuthal components of the magnetic field for frozen values of the plasma velocity and thermodynamic parameters (density and pressure); and the ``viscosity'' module (4) which computes the viscous dissipation in the disk.

{\bf (1)} In the hydrodynamic module, the ideal MHD equations are integrated numerically using an explicit conservative Godunov-type numerical scheme. In our numerical code, the dynamical variables are determined in the cells, while the vector-potential of the magnetic field, $A_\phi$, is determined on the nodes. To calculate the fluxes between the cells, we use the previously described approximate Riemann solver.

For better spatial resolution, we perform the reconstruction of the primitive variables to the boundaries between the calculated cells. We use two types of limiters: a) the MINMOD limiter \citep{Roe1986}, and b) a limiter which reconstructs the function with 3rd order accuracy (which is similar to the limiter developed by \citealt{Koren1993}). Integration of the equations in time is performed with a two-step Runge-Kutta method.

To guarantee a divergence-free field in the 2.5D version of the code, the $\phi$-component of the magnetic field vector-potential $A_\phi$ (or magnetic flux function $\Psi = r A_\phi$) is calculated at each timestep and then used to obtain the poloidal components of the magnetic field ($B_r$, $B_z$) in a divergence-free form \citep{Toth2000}. In other words, the condition $\nabla \cdot {\bf B} = 0$ is guaranteed to be satisfied to within machine accuracy. The calculation of $A_\phi$ is performed using either the algorithm proposed by \citet{BalsaraSpicer1999} or the modified form proposed by \citet{GardinerStone2005}.

{\bf (2)} In the module where the diffusion of the poloidal magnetic field is calculated, we numerically integrate the equation for the $\phi$-component of magnetic field vector-potential (or magnetic flux function $\Psi$), using an explicit approximation.

{\bf (3)} In the module where the diffusion of the azimuthal component of the magnetic field and {\bf (4)} the viscous stresses are calculated, we add diffusive terms which approximate the corresponding spatial derivatives to the fluxes calculated by the ideal MHD module.

The ideal and diffusivity modules are verified separately using a number of standard tests. The setup and results of these tests are discussed in Sec. \ref{sec:tests}.

\subsubsection{Ideal MHD module}
The ideal MHD equations in the axisymmetric cylindrical geometry are derived from the full set of 3D MHD equations by requring that $\partial/\partial{\phi}=0$ (see Sec.  \ref{subsec:geom 3D}). The equations are written in a form which does not have singularities on the right hand side. For the $r-$ and $\phi-$components of the equation of motion and $\phi-$component of the induction equation, we substitute the operator $\dst \frac{1}{r} \parti{}{r} r$, representing divergence operator in cylindrical coordinates. As a result, the $\phi-$component of the equations of motion takes on the form of angular momentum conservation while the $\phi-$component of the induction equation takes on its typical form.

In discretizing the cylindrical MHD equations, we take into account the fact that near the symmetry axis (where $r \to 0$), different variables have different asymptotic behavior. For example, the values $\rho, p, v_z, B_z \to \const$ as $r \to 0$ whereas the values $v_r, v_\phi, B_r, B_\phi \to 0$ scale with $r$. Analogous behavior is shown by the fluxes corresponding to these variables.  In particular, in the approximation of the $\phi-$component of the equation of motion (the angular momentum equation), we multiply the fourth equation (Eq. \ref{eq:momeqn_phi}) by $r^2$ and integrate it along $r$ and $z$ in the limits of the calculated cell $(i, j)$. The equation is then rewritten in the form $v_\phi = a_{i,j}r$ and $G = \rho v_\phi v_r - \dfrac{B_\phi B_r}{4\pi} = A_{i,j}r $, where $a_{i,j}$ and $A_{i,j}$ weakly depend on $r$. Here we have dropped the indices on $F$ and $G$.

We have
$$\prodi{}{t} \int a_{i,j} r^3 dr + \Delta_r(\cdot) + \Delta_z \int A_{i,j} r^3 dr = 0.$$
Next 
\begin{align*}
&\int a_{i,j} r^3 dr = \\
&\left< a_{i,j} \right> \dfrac{r_{i-1/2}^2 + r_{i+1/2}^2}{2} \dfrac{r_{i-1/2} + r_{i+1/2}}{2} (r_{i+1/2} - r_{i-1/2}) = \\
&v_{\phi,i,j} \left< r^2 \right>_i \Delta r_i ~, 
\end{align*}
where
\begin{align} 
v_{\phi,i,j} = & \left< a_{i,j} \right> \dfrac{r_{i-1/2} + r_{i+1/2}}{2}, \nonumber \\
\left< r^2 \right>_i = & \dfrac{r_{i-1/2}^2 + r_{i+1/2}^2}{2}, \nonumber\\
\Delta r_i = & r_{i+1/2} - r_{i-1/2}.
\end{align}
Analogously,
$$\int A_{i,j \pm 1/2} r^3 dr = G_{i,j \pm 1/2} \left< r^2 \right>_i \Delta r_i .$$
Thus, the angular momentum conservation equation for the cell $(i,j)$ has the form:
\begin{align}
\prodi{(\rho v_\phi)_{i,j}}{t} + \dfrac{r_{i+1/2}^2 F_{i+1/2,j} - r_{i-1/2}^2 F_{i-1/2,j}}{\left<r^2\right>_i \Delta r_i} \nonumber\\
+ \dfrac{G_{i,j+1/2} - G_{i,j-1/2}}{\Delta z_j} = 0.
\end{align}
It is clear that this approximation of the operator $\dst \frac{1}{r^2} \parti{}{r} r^2$ and more precisely of the multiplier $\left< r^2 \right>_i$, significantly differs from analogous terms (for example, of $r_i^2$) only in the vicinity of the symmetry axis.

The flux densities $F_{i \pm 1/2,j}$ and $G_{i,j \pm 1/2}$, which are used in Eq. \ref{eq:momeqn_phi}, are calculated using the 1D algorithm for the approximate solution of the Riemann problem.

\subsubsection{Viscosity and diffusivity} \label{subsec:viscosity and diffusivity}

Here we discuss the treatment of viscosity and diffusivity. We use equations of motion in cylindrical coordinates in the form:
$$\parti{\rho v_r}{t} + \parti{{\cal T}_{rr}}{r} + \parti{{\cal T}_{rz}}{z}  = \dfrac{{\cal T}_{\phi \phi} - {\cal T}_{rr}}{r} + \rho g_r,$$
\begin{equation}
\label{3-0}
\parti{\rho v_\phi}{t}
+ \frac{1}{r^2} \parti{r^2 {\cal T}_{r \phi}}{r} + \parti{{\cal
T}_{z \phi}}{z} = 0,
\end{equation}
$$
\parti{\rho v_z}{t} + \frac{1}{r} \parti{r {\cal T}_{rz}}{r}  + \parti{{\cal T}_{zz}}{z} = \rho g_z.
$$
The stress tensor ${\cal T}_{ik}=T_{ik}+\tau_{ik}$ consists of an ideal part $T_{ik}$:
$$
T_{ik}=\rho v_i v_k + \left( p + \frac{B^2}{8 \pi} \right)
\delta_{ik} - \frac{B_i B_k}{4 \pi} ~,~~~~i,k = r,\phi,z ~,
$$
which is used to derive the momentum equations in Eqs. \ref{eq:momeqn_r}, \ref{eq:momeqn_phi} and \ref{eq:momeqn_z}, and a viscous part $\tau_{ik}$ which takes into account small-scale turbulent velocity and magnetic field fluctuations.

We assume that the stress due to the turbulent fluctuations can be represented in the same way as the collisional viscosity by substituting in the turbulent viscosity coefficient. The components of the tensor $\tau_{ik}$ in axisymmetric cylindrical coordinates are:
\begin{align}
& \tau_{r\phi} = -\mu r \frac{\partial \omega}{\partial r} ~,
\tau_{z \phi} = - \mu r \frac{\partial \omega}{\partial z}~,  \tau_{\phi \phi} = - 2 \mu \frac{v_r}{r} ~,\nonumber \\
& \tau_{r z} = - \mu \frac{\partial v_r}{\partial z}~,
 \tau_{r r} = - 2 \mu\frac{\partial v_r}{\partial r}~,
\end{align}
Here $\omega = v_{\phi} /r $ is the angular velocity of the plasma and $\mu$ is the dynamical turbulent viscosity. In 3D, we only consider the $r\phi$ and $z\phi$ components of the stress tensor as these are expected to be dominant; similarly, in 2D polar coordinates, we only consider the $r\phi$ component.

Separating out the viscous stress in the $\phi$ component of the momentum equation gives
\begin{align}
\frac{\partial (\rho v_{\phi})}{\partial t} + \frac{1}{r^3}
\frac{\partial( r^3 T_{r \phi})}{\partial r} + \frac{1}{r \sin^2
\theta} \frac{\partial (\sin^2 \theta T_{z
\phi})}{\partial \theta} = \nonumber \\
 \frac{1}{r^3} \frac{\partial}{\partial r} \left( \nu \rho r^4
\sin \theta \frac{\partial \omega}{\partial r} \right) +
\frac{1}{r \sin^2 \theta} \frac{\partial}{\partial \theta} \left(
\nu \rho \sin^3 \theta \frac{\partial \omega}{\partial \theta}
\right),
\end{align}
where $T_{r \phi}$ and $T_{z \phi}$ are components of the inviscid part of the stress tensor.

We assume that the plasma has a finite magnetic diffusivity. That is, the matter may diffuse across the field lines. We also assume that the finite diffusivity of the plasma is due to the small-scale turbulent fluctuations of the velocity and the magnetic field. The induction equation averaged over the small-scale fluctuations has the form
\begin{equation}
\label{4-1}
\parti{\bf B}{t} - {\bf \nabla}\times ({\bf v}\times{\B})
+ {\bf \nabla}\times {\E}^\dagger =0~.
\end{equation}
Here, $\mathbf{v}$ and $\mathbf{B}$ are the averaged velocity and magnetic fields, and ${\bf E}^\dagger= - \left< {\bf
v}'\times{\B}' \right>$  is electromotive force connected with the fluctuating fields.

Since the turbulent electromotive force ${\bf E}^\dagger$ is connected with small-scale fluctuations, it is reasonable to suppose that it has a simple relation to the ordered magnetic field $\B$. If we neglect the magnetic dynamo $\alpha$-effect \citep{Moffat1978}, then $\left< {\bf v}'\times{\bf B}' \right> = -\eta {\bf \nabla}\times {\bf B}$, where $\eta$ is the coefficient of turbulent magnetic diffusivity. Equation (\ref{4-1}) now takes the form
\begin{equation}
\label{4-2}
\parti{\bf B}{t} - {\bf \nabla}\times ({\bf v} \times {\bf B})
+ {\bf \nabla} \times\left(\eta {\bf \nabla}\times {\bf B} \right) = 0~.
\end{equation}
We should note that the term for ${\E}^\dagger$ formally coincides with Ohm's law
$$
{\bf J} = \frac{c}{4\pi} {\bf \nabla}\times {\B} = \frac{c^2}{4\pi \eta} {\E}^\dagger~.
$$
The coefficient of turbulent electric conductivity $\sigma = c^2 /4\pi \eta$.


To calculate the evolution of the poloidal magnetic field it is useful to calculate the $\phi$-component
of the vector-potential $\mathbf{A}$ or magnetic flux function $\Psi = r A_\phi$. Due to the assumed axisymmetry,
\begin{equation}
\label{4-3}
B_r = - \dfrac{1}{r} \parti{\Psi}{z},~~~~B_z = \dfrac{1}{r} \parti{\Psi}{r}.
\end{equation}
Substituting  $\B = {\bf \nabla}\times {\bf A}$ into the induction equation gives the equation for the $\phi$ component
of the vector-potential
\begin{equation}
\label{4-4}
\parti{\Psi}{t} - \eta \left( r \parti{}{r} \frac{1}{r} \parti{\Psi}{r}
+ \partii{\Psi}{z} \right) = r ({\bf v}\times{\B})_{\phi}~.
\end{equation}

The azimuthal component of the induction equation gives
\begin{equation}
\parti{B_\phi}{t} -
\parti{}{r} \left( \frac{\eta}{r} \parti{r B_\phi}{r} \right)
+ \parti{}{z} \left( \eta \parti{B_\phi}{z} \right) = \left( { 
\nabla} \times (\mathbf{v} \times \mathbf{B}) \right)_\phi. \label{4-5}
\end{equation}

The viscosity and magnetic diffusivity lead to dissipation of the kinetic and magnetic energies, conversion into thermal energy and a corresponding increase of the entropy. In our simulations we have neglected viscous and Ohmic heating. We have also neglected radiative cooling. Thus, the main effect of the viscous terms is the transport of angular momentum outward which allows matter to accrete inward to the disk-magnetosphere boundary. The main effect of the diffusion terms is the transport of matter through magnetic field lines.

The dissipative terms $\tau_{r \phi}, \tau_{z \phi}$ and, $\E^\dagger$ are incorporated into the numerical algorithm by way of a flux correction in the finite-difference equations. The modified fluxes $F_4,~G_4$ are included in the angular momentum conservation equations, and the modified fluxes $F_7,~G_7$ (components of the poloidal electric field) are included into the $\phi-$component of the induction equation. The approximation of the viscous stresses takes the form:
\begin{align*}
F_{4,i+1/2,j} &=& F_{4,i+1/2,j}^{id} - \mu_{i+1/2,j} r_{i+1/2}
\dfrac{\omega_{i+1,j}-\omega_{i,j}}{\Delta r_{i+1/2}}, \\
G_{4,i,j+1/2} &=& G_{4,i,j+1/2}^{id} - \mu_{i,j+1/2} r_i
\dfrac{\omega_{i,j+1}-\omega_{i,j}}{\Delta z_{j+1/2}}.
\end{align*}
The approximation of the poloidal electric field has the form:
\begin{align*}
F_{7,i+1/2,j} &=& F_{7,i+1/2,j}^{id} - \eta_{i+1/2,j}
\dfrac{r_{i+1} B_{\phi,i+1,j} - r_i B_{\phi,i,j} }{(r \Delta
r)_{i+1/2}},
\\
G_{7,i,j+1/2} &=& G_{7,i,j+1/2}^{id} - \eta_{i,j+1/2} r_i \dfrac{
B_{\phi,i,j+1} - B_{\phi,i,j} }{\Delta z_{j+1/2}}.
\end{align*}
Here, $F_{7,i+1/2,j}$ and $G_{7,i,j+1/2}$ are the $z-$ and $r-$components of electric field, and $F_{7,i+1/2,j}^{id}, G_{7,i,j+1/2}^{id}$ are the same components calculated in the approximation of the ideal plasma conductivity using the algorithm described above, $(r \Delta r)_{i+1/2} = (r_{i+1}^2 - r_i^2)/2$.

The coefficients of the dynamical viscosity at the boundaries between calculated cells are calculated according to $\dst \mu_{i+1/2,j} = \frac{2 \mu_{i,j} \mu_{i+1,j}}{\mu_{i,j} + \mu_{i+1,j}}$. Here, $\mu_{i,j}$ and $\mu_{i+1,j}$ are the coefficients of the dynamical viscosity in the grid cells.  Analogously, we calculate $\mu_{i,j+1/2}$ and the coefficient of the dynamical viscosity at the boundaries between cells.

The azimuthal component of the electric field, responsible for the evolution of the magnetic flux function $\Psi$, is approximated using a special method. The azimuthal component of the turbulent electric field $E_\phi^\dagger$ is presented in the form
$$
- \eta \left( r \parti{}{r} \frac{1}{r} \parti{\Psi}{r}
+ \partii{\Psi}{z} \right)
$$
and approximated by
\begin{equation}
\begin{split}
E_{\phi,i+1/2,j+1/2}^* = \eta_{i+1/2,j+1/2} \bigg[
\dfrac{r_{i+1/2}}{\Delta r_{i+1/2}} \\
 \left(
\dfrac{\Psi_{i+3/2,j+1/2}-\Psi_{i+1/2,j+1/2}}{(r \Delta r)_{i+1}}
\right.\\~\left.- \dfrac{\Psi_{i+1/2,j+1/2}-\Psi_{i-1/2,j+1/2}}{(r
\Delta r)_{i}} \right)
 \\ +
\dfrac{1}{\Delta z_{j+1/2}} \left(
\dfrac{\Psi_{i+1/2,j+3/2}-\Psi_{i+1/2,j+1/2}}{\Delta z_{j+1}}
\right.\\~\left.-
\dfrac{\Psi_{i+1/2,j+1/2}-\Psi_{i+1/2,j-1/2}}{\Delta z_j} \right)
\bigg].
\end{split}
\end{equation}


A corresponding term is added to Eq. \ref{4-4}, which takes on the form
\begin{equation}
\begin{split}
\frac{\Psi_{i+1/2,j+1/2}^{n+1} - \Psi_{i+1/2,j+1/2}^n}{\Delta t} -
\eta_{i+1/2,j+1/2} \bigg[ \dfrac{r_{i+1/2}}{\Delta r_{i+1/2}} \\
\left( \dfrac{\Psi_{i+3/2,j+1/2}-\Psi_{i+1/2,j+1/2}}{(r \Delta
r)_{i+1}} \right.\\~\left. -
\dfrac{\Psi_{i+1/2,j+1/2}-\Psi_{i-1/2,j+1/2}}{(r \Delta r)_{i}}
\right) \\ + \dfrac{1}{\Delta z_{j+1/2}} \left(
\dfrac{\Psi_{i+1/2,j+3/2}-\Psi_{i+1/2,j+1/2}}{\Delta z_{j+1}}
\right.\\~\left.-
\dfrac{\Psi_{i+1/2,j+1/2}-\Psi_{i+1/2,j-1/2}}{\Delta z_j} \right)
\bigg] = - r_{i+1/2} E_{\phi,i+1/2,j+1/2}^{id}.
\end{split}
\end{equation}
The values of $E_{\phi,i+1/2,j+1/2}^{id}$ on the right hand side are calculated using Eqs. \ref{2-2} and \ref{2-3}.  At the first step of the time integration, we use a time-step of $\Delta t/2$ and calculate the values of $\Psi_{i+1/2,j+1/2}^{n+1/2}$.

\subsection{3D cylindrical geometry} \label{subsec:geom 3D}

We have also developed a fully three dimensional version of the code in cylindrical coordinates. To calculate the fluxes between cells, we use the HLLD method as in the 2.5D code \citep{MiyoshiKusano2005}. To correct the fluxes, we use the MINMOD flux corrector, which is applied on the primitive variables $\rho, s, v_r, v_\phi, v_z, B_r, B_\phi, B_z$. To integrate the equations in time, we use the two-step Runge-Kutta method.

The main differences between the 2D and 3D codes lie in the algorithm which ensures that the magnetic field is divergence-free. Namely, the divergence-free condition is guaranteed by calculating the components of the magnetic field on the sides of cells from the induction equation:
$$\parti{\B}{t} + \rot \E = 0.$$
The components of the electric field are calculated at the edges of the cell as in the axisymmetric case. For example:
\begin{eqnarray*} 
E_{z,i+1/2,j+1/2,k} = \frac{1}{4} \left( E_{z,i,j+1/2,k} +
E_{z,i+1,j+1/2,k} \right.\\
~\left. + ~E_{z,i+1/2,j,k} + E_{z,i+1/2,j+1,k} \right),
\end{eqnarray*}
where the values on the right hand side are the corresponding fluxes calculated by the Riemann solver: $E_{z,i,j+1/2,k}$ is determined between cells $(i,j,k)$ and $i,j+1,k$ and so on.

After calculating the electric field components, the magnetic field components at the sides of cells (for example, $B_r$ at $(i+1/2,j,k)$) are calculated using the formula
\begin{align*}
\frac{\bar{B}_{r,i+1/2,j,k}^{n+1} -
\bar{B}_{r,i+1/2,j,k}^n}{\Delta t} + \frac{E_{z,i+1/2,j+1/2,k}^n -
E_{z,i+1/2,j-1/2,k}^n}{r_{i+1/2} \Delta \phi} \\ +
\frac{E_{\phi,i+1/2,j,k-1/2}^n - E_{\phi,i+1/2,j,k+1/2}^n}{\Delta
z} = 0.
\end{align*}

The other components $\bar{B}_{\phi,i,j+1/2,k}$ and $\bar{B}_{z,i,j,k+1/2}$ are calculated in a similar fashion. This ensures that the magnetic field $(\bar{B}_{r,i+1/2,j,k}, \bar{B}_{\phi,i,j+1/2,k}, \bar{B}_{z,i,j,k+1/2})$ is divergence-free (so long as the field is initially divergence-free at the beginning of the simulation). As a next step, the magnetic field in the cells is calculated from the values on the cell boundaries:
\begin{align*}
&B_{r,i,j,k} = \frac{1}{2} \left( \bar{B}_{r,i-1/2,j,k} +
\bar{B}_{r,i+1/2,j,k} \right),
\\
&B_{\phi,i,j,k} = \frac{1}{2} \left( \bar{B}_{\phi,i,j-1/2,k} +
\bar{B}_{\phi,i,j+1/2,k} \right),
\\
&B_{z,i,j,k} = \frac{1}{2} \left( \bar{B}_{z,i,j,k-1/2} +
\bar{B}_{z,i,j,k+1/2} \right).
\end{align*}

\subsection{2D polar geometry}
\label{subsec:geom polar coord}

To describe the interaction of the accretion disk with a planet in the presence of a magnetic field, we use a model which describes the processes in terms of surface variables (density and pressure). The equations for this model are obtained by integrating the three-dimensional MHD equations along the $z-$direction which is perpendicular to the disk plane. For simplicity, we assume that the $z-$components of the velocity and the magnetic field are zero and there is no field outside the disk.

The governing equations are derived from the ideal MHD equations (Eq. \ref{eq:full-3d}) by integrating in the vertical direction:
\begin{equation}
\parti{\Sigma}{t} + \frac{1}{r} \parti{r \Sigma v_r}{r} + \frac{1}{r} \parti{\Sigma v_\phi}{\phi} =0,
\label{8-1}
\end{equation}
\begin{align}
&\dst \parti{\Sigma v_r}{t} + \parti{}{r} \left( \Sigma v_r^2 + \Pi + \frac{b_r^2 + b_\phi^2}{8 \pi} -
\frac{b_r^2}{4\pi} \right) + \nonumber\\
& \frac{1}{r}
\parti{}{\phi} \left( \Sigma v_r v_\phi - \frac{b_r
b_\phi}{4\pi} \right) = \frac{ \Sigma (v_\phi^2-v_r^2) - \frac{b_\phi^2-b_r^2}{4 \pi}}{r}- \Sigma \frac{GM_*}{r^2} +
\Sigma f_r, \label{8-7}
\end{align}
\begin{align}
&\parti{\Sigma v_\phi}{t} + \frac{1}{r^2} \parti{}{r}r^2 \left(\Sigma v_r v_\phi - \frac{b_r b_\phi}{4 \pi} \right) +
\nonumber\\
&\frac{1}{r} \parti{}{\phi} \left(\Sigma v_\phi^2 + \Pi + \frac{b_r^2 + b_\phi^2}{8 \pi} - \frac{b_\phi^2}{4 \pi}
\right) = \Sigma f_\phi. \label{8-8}
\end{align}
\begin{equation}
\parti{\Sigma S}{t} + \frac{1}{r} \parti{r \Sigma S v_r}{r} + \frac{1}{r} \parti{\Sigma S v_\phi}{\phi} = 0 ~.
\label{8-4}
\end{equation}
Here $\Sigma = \int \rho dz$ is the surface density; $\Pi = \int p dz$ is the surface pressure; $\mathbf{b} = (b_r, b_\phi, b_z)$ is the $z$-averaged analog of the mangetic field; $S = \Pi/\Sigma^{\gamma}$ is an analog of the entropy; $\Sigma f_r$, and $\Sigma f_\phi$ are the components of the force per unit surface area acting from the planet on the disk matter. These resemble the standard form of the 2D equations where the $z-$derivative in the 3D equations is set to 0.

The induction equations take on the form
\begin{equation}
\parti{\sqrt{h_m} b_r}{t} + \frac{1}{r} \parti{}{\phi} \sqrt{h_m} (v_\phi b_r - v_r b_\phi) = 0,
\label{8-9}
\end{equation}
\begin{equation}
\parti{\sqrt{h_m} b_\phi}{t} + \parti{}{r} \sqrt{h_m} (v_r b_\phi - v_\phi b_r) = 0.
\label{8-10}
\end{equation}
Here, $h_m$ is the ``magnetic thickness'' of the disk, which is the characteristic scale for the magnetic field distribution in $z-$direction. This scale is not identical to the disk height $h$ which characterizes the $z-$distribution of the density and pressure in the disk. We take $h_m$ to be the same across both equations. Since the magnetic field in the disk is unlikely to change sign in the vertical direction, the approximation seems to be reasonable. The next suggestion which helps to close the system of equations is in the dependence of the magnetic thickness of the disk $h_m(r)$ on the radius. In particular, if we assume $h_m$ is constant, then the induction equations will also take the standard form (in the same sense that $\parti{}{z}=0$). To calculate the magnetic field and volumetric density and pressure from the $z-$integrated values, we assume a disk thickness at some characteristic radius and back out the volumetric terms.

\section{Tests of the Code}
\label{sec:tests}

We have performed multiple tests of the code across different grid geometries. Below we show two examples of such tests. In the first, we test the ideal MHD module (with the viscosity and diffusivity modules switched off).  In the second, we test the diffusion module separately.

\subsection{Test of the ideal MHD module}
\subsubsection{The 2D rotor problem}

\begin{figure}
\centering
\includegraphics[width=\columnwidth]{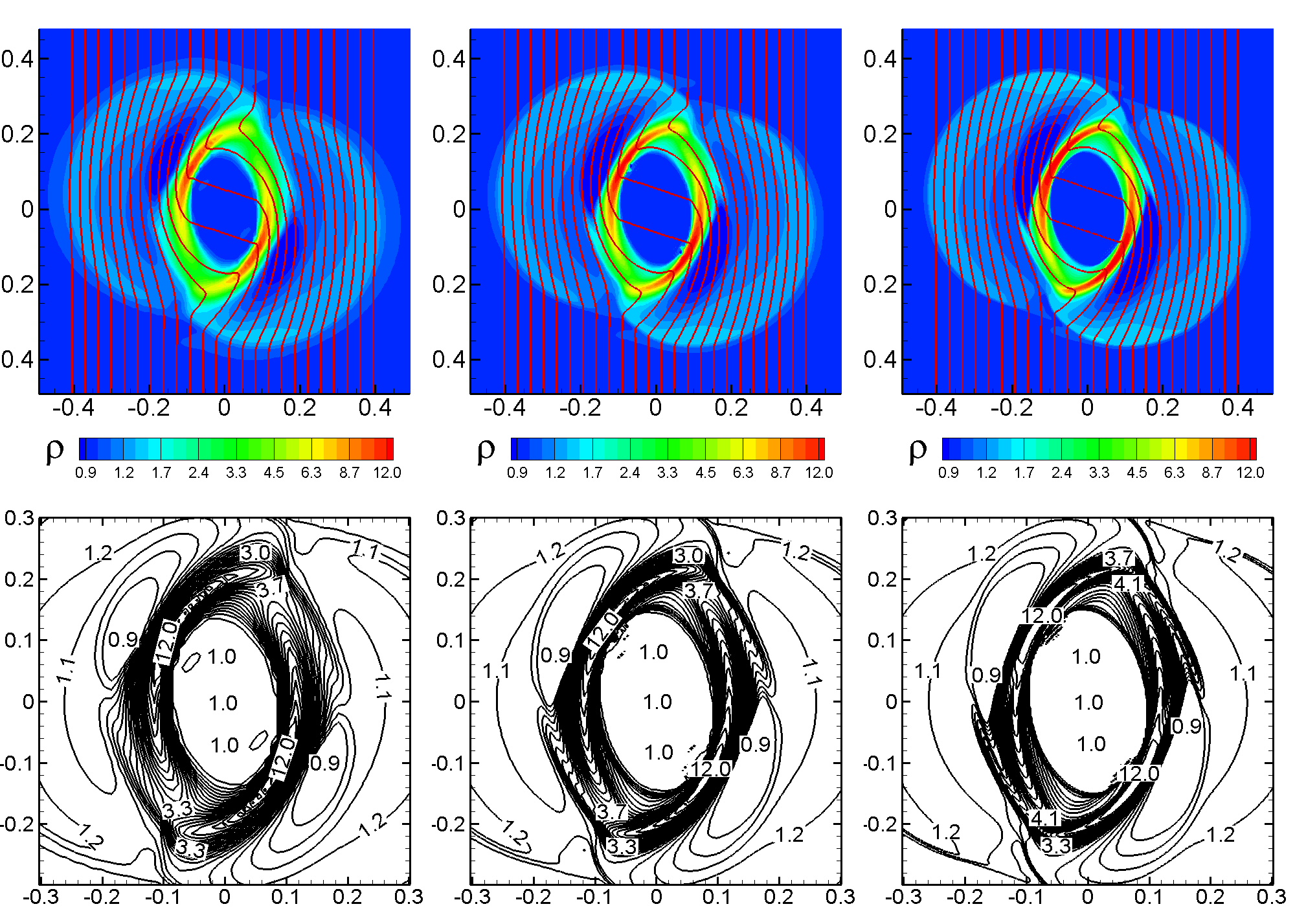}
\caption{The test of the ideal MHD module of the code with the ``rotor problem" at different grid resolutions. {\it Top Panels:} Density distribution (color background) and field lines (solid lines) with grid resolution $100\times 100$ (left panel), $200\times 200$ (middle panel) and $400\times 400$ (right panel).  {\it Bottom Panels:} The density contour lines.} \label{ideal-6}
\end{figure}

The first test is the standard ``rotor problem" in Cartesian coordinates. This test has been used by a number of authors for testing MHD solvers which use the energy equation \citep{BalsaraSpicer1999,Toth2000}. We use this test to check the ideal MHD module of our code (with viscosity and diffusivity switched off).

The ideal MHD equations are solved on a regular Cartesian grid in the region $-0.5<x<0.5$, $-0.5<y<0.5$ with grid sizes $\Delta x = \Delta y = 1/N$, where $N=100, 200, 400$ for the three different tests. At the beginning of the simulations, $t=0$, the pressure in the region is constant, $p=1$, and the magnetic field is homogeneous, $B_x=0$, $B_y=5$. In the center, there is a circle of higher-density matter ($\rho_0=10$) with radius $r_0=0.1$ (where the radius is $r=\sqrt{x^2+y^2}$). The matter in the inner circle initially rotates as a solid body with angular velocity $\omega_0=20$. For $r>r_1=0.115$, the density is $\rho_1=1$ and the matter is at rest. In the ring in between these two regions $r_0<r<r_1$, the density and velocity are linearly interpolated between the values at $r=r_0$ and $r=r_1$.

\begin{figure}
\centering
\includegraphics[width=0.45\textwidth]{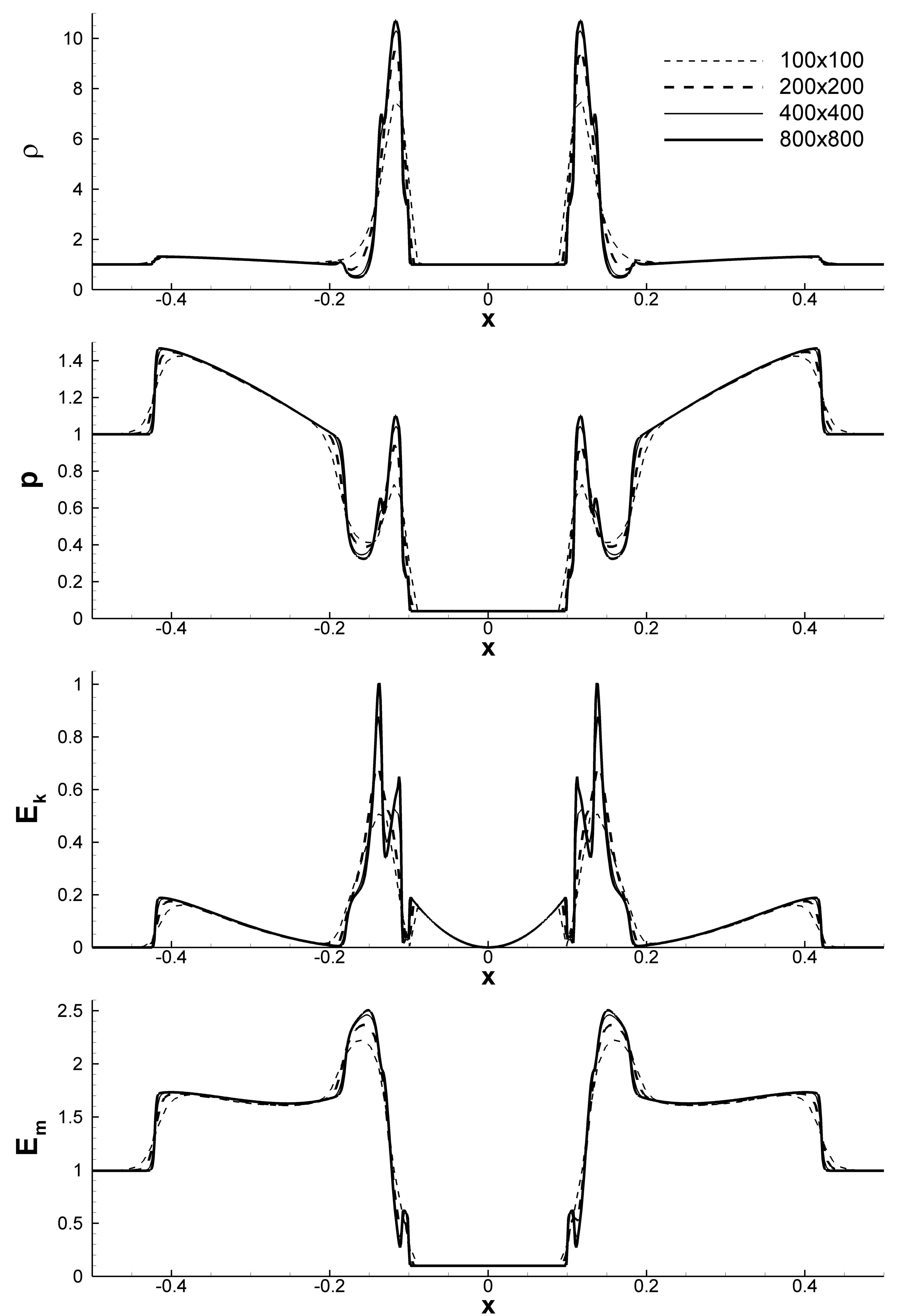}
\caption{Slices of the density $\rho$, pressure $p$, kinetic energy, $E_k$ and magnetic energy $E_m$ along a horizontal line at the center of the rotor at time $t = 0.15$. \label{fig:rotor_cuts}}
\end{figure}

The equations of ideal adiabatic MHD are solved with the previously described Godunov-type scheme. The timesteps are calculated from the condition $\Delta t=0.4 \Delta t_{CFL}$ where $\Delta t_{CFL}$ is the maximum timestep allowed by the Courant-Friedrichs-Levy (CFL) condition.  The results of the simulations at $t = 0.15$ are shown in \fig{ideal-6}. Evidently, the density and the field line distribution is very similar in all three of our cases. The bottom panels of \fig{ideal-6} show selected streamlines with numbers which clearly show the similarity of the results across the various grid sizes. \fig{fig:rotor_cuts} shows slices along the $x$-axis for the density, pressure, kinetic energy, and magnetic energy for each of the three grid sizes in addition to a high resolution grid with $N = 800$. At the highest resolutions, the models are nearly indistinguishable and the convergence of the results is evident.

\subsubsection{2D Low-type analytical solutions} \label{subsubsec:2Dlow}
\begin{figure}
\centering
\includegraphics[width=2.8in]{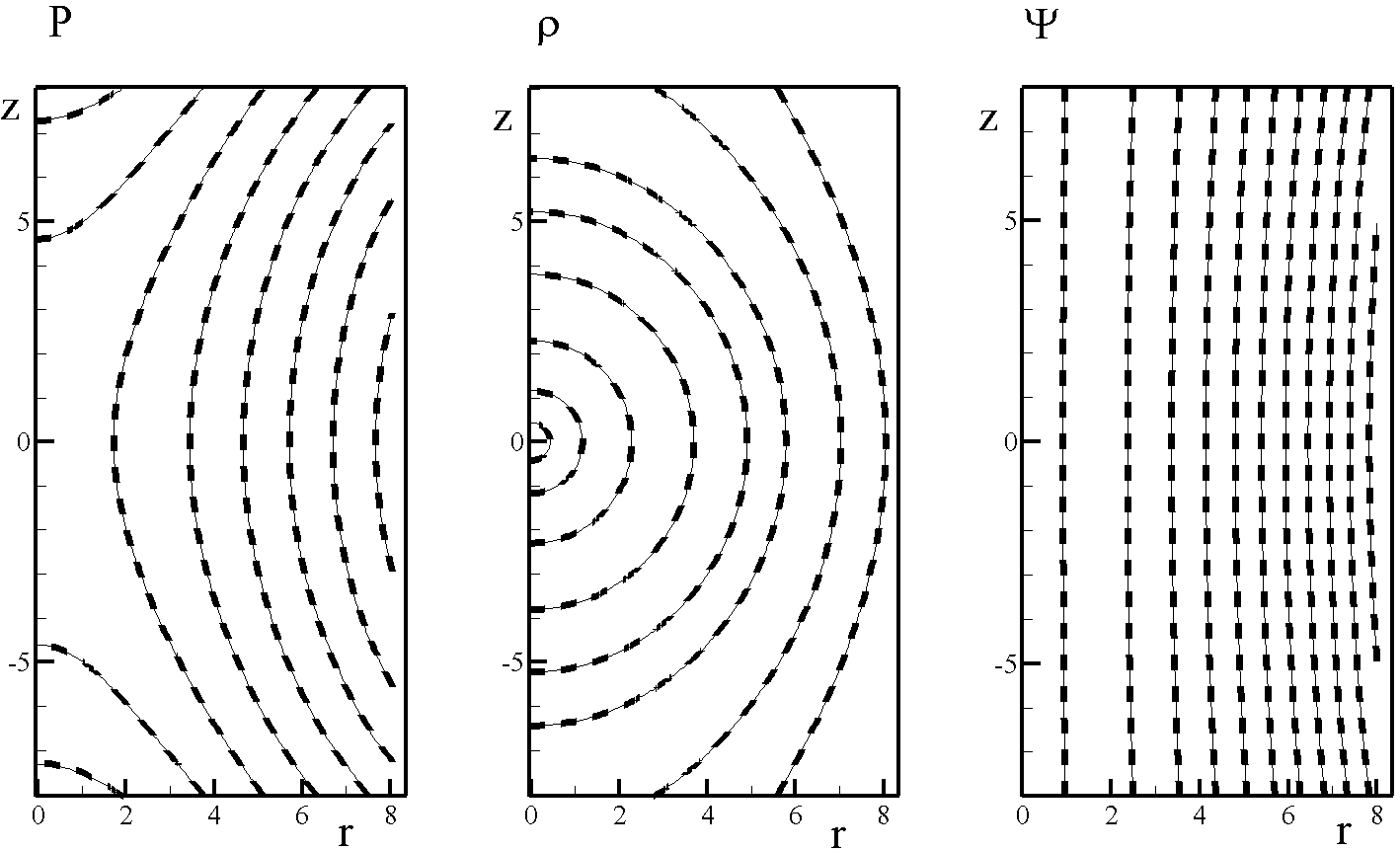}
\caption{The test of the ideal MHD module with the Low-type solution using a homogeneous grid with $N_r, N_z = 80 \times 160$. From left to right: contours of the pressure, density and magnetic flux functions at the moment of time when the scaling parameter is $a=6$. In all of the plots, the solid line shows the numerical solution while the dashed line shows the analytical Low-type solution.}
\label{lou-un-a6}
\end{figure}

\begin{figure}
\centering
\includegraphics[width=2.8in]{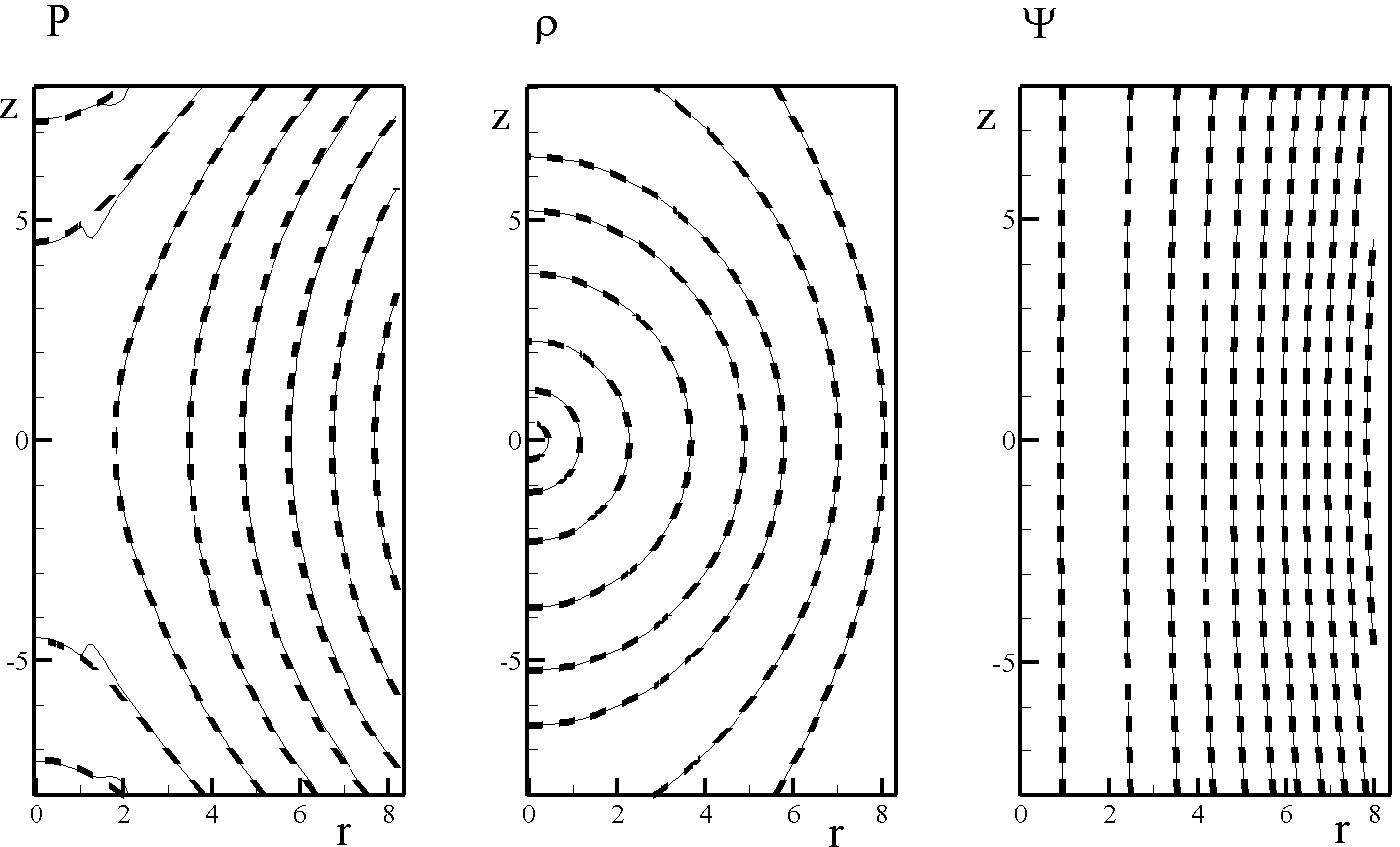}
\caption{The same as in Fig. \ref{lou-un-a6}, but using an inhomogeneous grid with $N_r, N_z = 40 \times 80$.} \label{lou-nun40-a6}
\end{figure}

\begin{figure}
\centering
\includegraphics[width=3in]{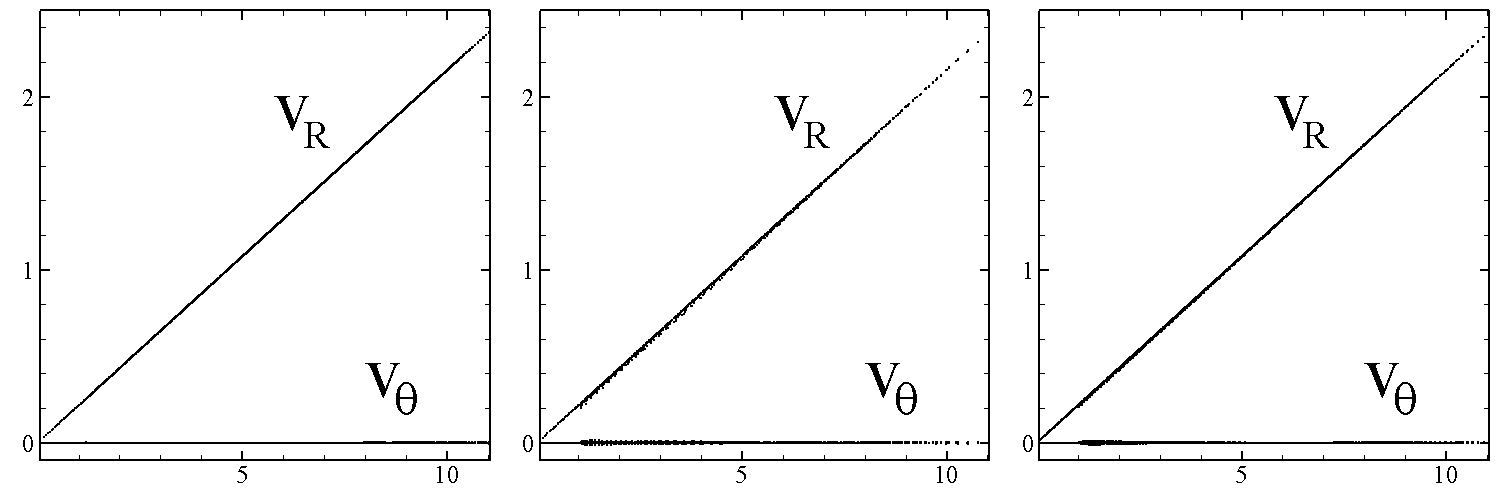}
\caption{The test of the ideal MHD module using the Low-type solution on a homogeneous grid with $N_r, N_z = 80 \times 160$. The dependence of the radial component of velocities and components perpendicular to the radius on the spherical radius for all grid cells.}
\label{lou-v-a6}
\end{figure}

\begin{figure}
\centering
\includegraphics[width=6cm]{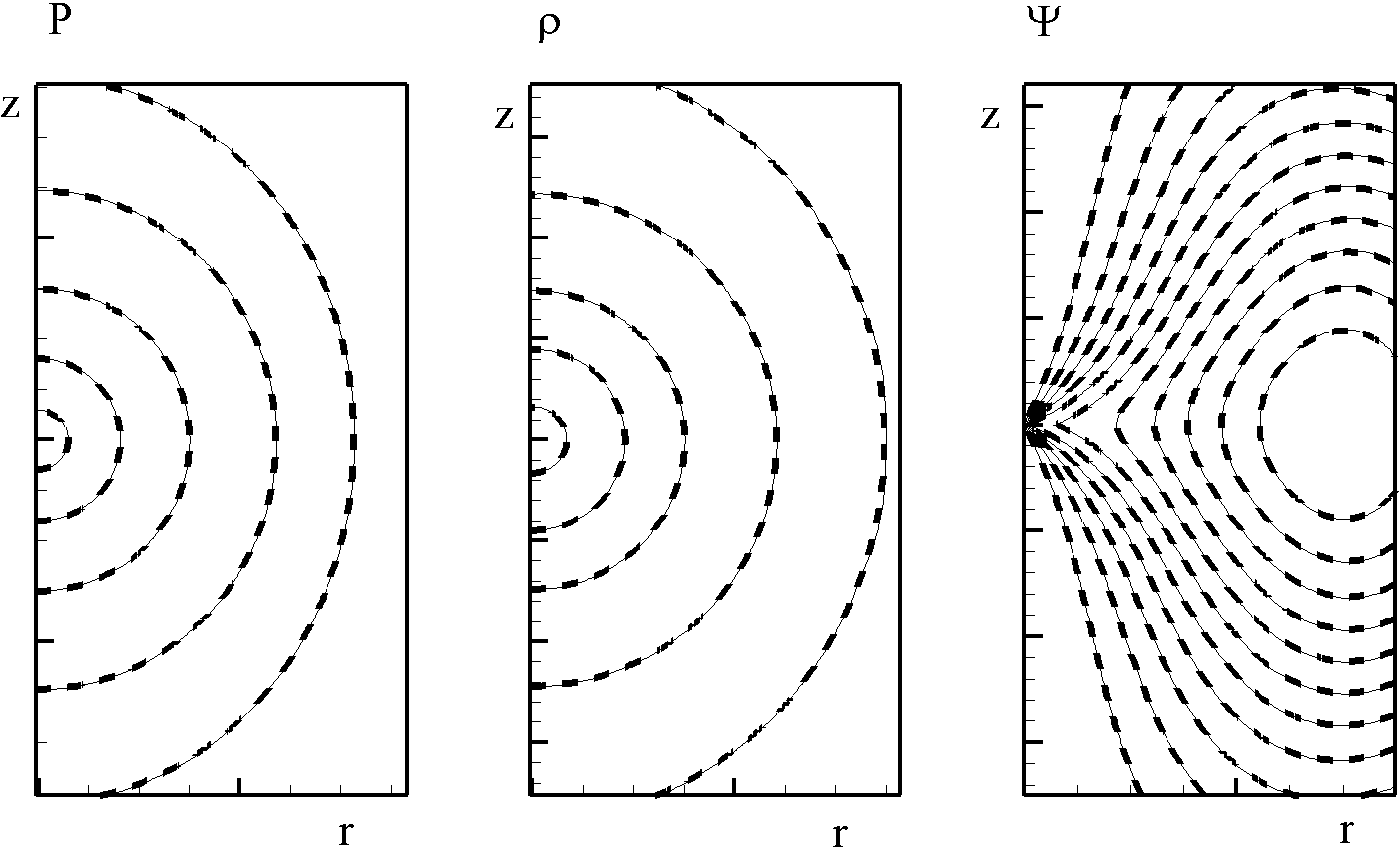}
\caption{The test of the ideal MHD module with solution of \cite{Stone1992} on homogeneous grid with $N_r, N_z = 70 \times 140$. From left to right: pressure, density and magnetic flux function at the moment of time when the scaling parameter is $a=6$. In all plots, the solid line shows the numerical solution, dashed line --– analytical solution.} \label{stone-a6}
\end{figure}

To test the ideal MHD module, we also compare our numerical calculations with analytic solutions found by \cite{Low1984}.  One of them has been used in \cite{Stone1992} for testing the corresponding module of the ZEUS code. It is convenient to write these solutions in spherical coordinates (as we will do below), and then to convert them to cylindrical coordinates.

These solutions are self-similar and assume that the magnetic flux function $\Psi$ has a dependence on the time and radial coordinate of the form $\xi = R/a(t)$. Here $a$ is a time-dependent scaling parameter and $\xi$ is the self-similarity radius. The velocity has only a radial component (in the spherical coordinate system) which depends linearly on the radius. The function $a(t)$ satisfies the equation $\dst \frac{1}{2} \left( \prodi{a}{t} \right)^2 + \frac{\alpha}{a} = \beta$, and the magnetic flux function satisfies the Grad-Shafranov equation. The other variables are connected with $a(t)$ and $\Psi$ and their derivatives by algebraic relationships. The first solution proposed by \cite{Lou1998} and used for testing does not have an azimuthal component of the magnetic field. The second solution used in \cite{Stone1992} has a non-zero azimuthal magnetic field $B_\phi$, but the motion of the plasma is inertial.

In the first solution, \cite{Lou1998} chose growing solutions of the function $a(t)$
\begin{equation}
\label{5-1}
\dst
\prodi{a}{t} = \sqrt{2 \left(\beta - \frac{\alpha}{a} \right)} ~.
\end{equation}
The magnetic flux function has the form
$$
\Psi = C \xi^2 (\xi^2 - \xi_0^2)^2 \sin^2 \theta ~.
$$
The poloidal components of the magnetic field are calculated along the known formulae
\begin{align*}
&B_R = \dfrac{2 C \cos \theta}{a^2} (\xi^2 - \xi_0^2)^2, \\
&B_\theta = - \dfrac{2 C \sin \theta}{a^2} (\xi^2-\xi_0^2) (3 \xi^2-\xi_0^2) ~.
\end{align*}
The initial pressure and density are
\begin{align*}
&p=\frac{C^2}{\pi a^4} \bigg[(5 \xi_0^2 - 7 \xi^2) (\xi^2 -
\xi_0^2) \xi^2 \sin^2 \theta - \xi_0^6 \bigg] (\xi^2 - \xi_0^2), \\
&\rho = \frac{2 C^2 \xi}{\pi a^3 (\alpha \xi + GM/\xi^2)} \bigg[7 (\xi^2 - \xi_0^2)^2 \xi^2 \sin^2 \theta + \xi_0^6 \bigg].
\end{align*}
And the radial velocity is given by:
$\dst v_R = \xi \sqrt{2 \left( \beta - \frac{\alpha}{a} \right)}.$

We take the following parameters for the problem: $GM=1,~C=1,~\alpha=1,~\beta=1,~\xi_0 = 6$, the adiabatic index is $\gamma=4/3$.

The solution is numerically computed over the region $0<r<8$, $-8 < z < 8$ with a rectangular region, $0<r<1$, $-1 < z < 1$ excised from the simulation. We use two grids: a homogeneous grid, with $N_r \times N_z = 80 \times 160$, and an inhomogeneous grid, $40 \times 80$. In both cases, the excised rectangular region is homogeneous, with dimensions $10 \times 10$, though values are not computed in this region. Thus, the cell sizes of the homogenous grid are $\Delta r =0.1$ and $\Delta z =0.1$. For the inhomogeneous grid, the grid size is increased in a geometrical progression with the power $q=1.05$. The integration time-step has been chosen automatically such that $\Delta t = 0.5 \Delta t_{CFL}$. The function $a(t)$ is determined by numerically integrating Eq. \ref{5-1} with the initial condition $a(0)=2$ at sufficiently fine grid resolution. As an initial condition, the analytical solution is used with the parameter $a=2$. For boundary conditions, we also use the analytical solution for all variables with the parameter $a$ corresponding to the moment of time. The simulations were performed up to the moment of time when $a=6$. \fig{lou-un-a6} and \fig{lou-nun40-a6} from left to right show the pressure, density and magnetic flux function at the moment of time when the scaling parameter is $a=6$. \fig{lou-un-a6} shows the result of the simulations on a homogeneous grid, while \fig{lou-nun40-a6} shows the results for the inhomogeneous grid.  In all of the plots, the solid line shows the numerical solution and the dashed line shows the analytical solution. \fig{lou-v-a6} shows the radial and transverse components of the velocities as a function of the spherical radius.  One can see that the velocity is directed along the radial direction and is proportional to the radius with the high accuracy.

\begin{figure}
\centering
\includegraphics[width=2.0in]{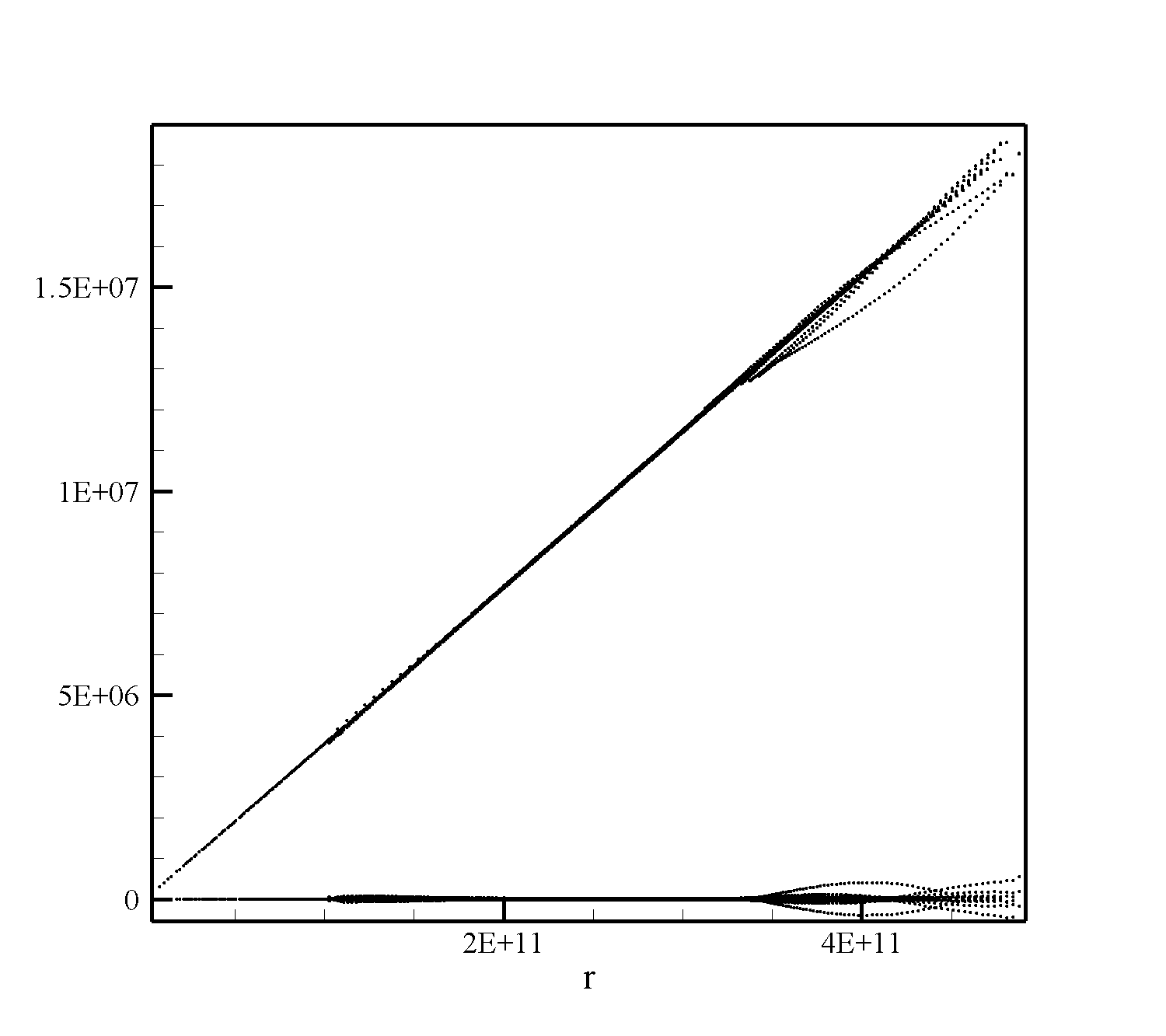}
\caption{The test of the ideal MHD module using the solution of \cite{Stone1992} on homogeneous grid with $N_r, N_z = 70 \times 140$. The dependence of the radial component of velocities and components perpendicular to the radius on the spherical radius for all grid cells.} \label{stone-v-a6}
\end{figure}

In the second solution \citep{Stone1992} the scaling factor linearly depends on time, $a(t) = t \sqrt{\eta}$.  In this case, the magnetic flux function has the form
$$
\Psi = \left \{
\begin{array}{ll}
\dst{A_0 \left(p_0 + \frac{\sin \lambda \xi}{\lambda \xi} - \cos \lambda \xi \right) \sin^2 \theta } &
\xi < \xi_c \\[0.3cm]
\dst{0}&
\xi > \xi_c~.
\end{array}
\right.
$$
The poloidal components of the magnetic field (in spherical coordinates) can be calculated using the analytic solution
$$
B_R = \left \{
\begin{array}{ll}
\dst{\dfrac{2 A_0 \cos \theta}{R^2} \left(p_0 + \frac{\sin \lambda
\xi}{\lambda \xi} - \cos \lambda \xi \right)} &
\xi < \xi_c \\[0.3cm]
\dst{0}& \xi > \xi_c~.
\end{array}
\right.
$$
$$
B_\theta = \left \{
\begin{array}{ll}
\dst{- \dfrac{\lambda A_0 \sin \theta}{a R} \left( \sin \lambda
\xi + \frac{\cos \lambda \xi}{ \lambda \xi} -\frac{\sin \lambda
\xi}{\lambda^2 \xi^2} \right)} &
\xi < \xi_c \\[0.3cm]
\dst{0}& \xi > \xi_c~.
\end{array}
\right.
$$
The azimuthal component of the magnetic field is
\begin{eqnarray}
B_\phi = \dfrac{\lambda \Psi}{a^2 R \sin \theta} = \nonumber \\
 \left \{
\begin{array}{ll} \dst{\frac{\lambda A_0 \sin \theta }{a R}
\left(p_0 + \frac{\sin\lambda \xi}{\lambda \xi} - \cos \lambda \xi
\right) } &
\xi < \xi_c \\[0.3cm]
\dst{0}& \xi > \xi_c~.
\end{array}
\right.
\end{eqnarray}
The density and pressure are given by
$$
\rho = \rho_s + \frac{A_0^2 p_0}{2 \pi GM R^3} (4 - \lambda^2 \xi^2)
\left(p_0 + \frac{\sin \lambda \xi}{\lambda \xi} - \cos \lambda \xi \right) \sin^2 \theta,
$$
$$
\rho_s = \left \{
\begin{array}{ll}
\dst{\left( \dfrac{GM}{\nu R} \right)^3} &
\xi < \xi_c \\[0.3cm]
\dst{\frac{7 d_0}{a^3} \left(\frac{R_0}{\xi} \right)^8}&
\xi > \xi_c~.
\end{array}
\right.
$$
$$
p = p_s + \frac{A_0^2 p_0}{4 \pi R^4} (2 - \lambda^2 \xi^2)
\left( p_0 + \frac{\sin \lambda \xi}{\lambda \xi} - \cos \lambda \xi \right) \sin^2 \theta,
$$
$$
p_s = \left \{
\begin{array}{ll}
\dst{\dfrac{GM \rho_s}{4 R}} &
\xi < \xi_c \\[0.3cm]
\dst{\dfrac{7 d_0 \eta R_0^2  }{6 a^4} \exp \left( \frac{2 GM}{3 \eta R_0^3} \left(\frac{R_0}{\xi}\right)^9 \right)}&
\xi
> \xi_c~.
\end{array}
\right.
$$
$$
d_0=10^8 m_p \exp \left( - \frac{2 GM}{3 \eta R_0^3} \right),~~~~ m_p= 1.673 \times 10^{-24}g.
$$
The radial velocity is simply $v_r = \xi \sqrt{\eta} = R/t$.

The parameters of the problem are assigned the following values: $\dst M = 2 \times 10^{33}{\rm g}, \eta = 5.24 \times 10^{-8} {\rm sec}^{-2},~~ \lambda = 5.54 \times 10^{-11}{\rm cm}^{-1},~~\xi_c = 1.104 \times 10^{11}{\rm cm},~~ \nu = 2.42 \times 10^{20} {\rm cm^3 sec^{-2} g^{-1/3}},~~A_0=1.5 \times 10^{21} {\rm G cm^2},~~ p_0 = \cos \lambda \xi_c - \frac{\sin \lambda \xi_c}{\lambda \xi_c} = 1.01327,~~R_0=10^{11}{\rm cm}$, the adiabatic index is $\gamma=4/3$.

The equations of ideal MHD are integrated in the region $0< r <3.5 \times 10^{11}{\rm cm}$, $-3.5 \times 10^{11}{\rm cm} < z < 3.5 \times 10^{11}{\rm cm}$, where an inner rectangle with size $0 < r < 10^{11}{\rm cm}$, $-10^{11}{\rm cm} < z < 10^{11}{\rm cm}$ has been excised from the region. We use the homogeneous grid $70 \times 140$, and the grid in the excised region is $20 \times 20$. Thus, the cell size is $\Delta r = 0.05 \times 10^{11}{\rm cm}$ and $\Delta z = 0.05 \times 10^{11}{\rm cm}$. The time-step is chosen automatically such that $\Delta t = 0.5 \Delta t_{CFL}$. As an initial condition, we take the analytical solution described above for $a=2$. We use the analytical solution for the boundary conditions, taking the parameter $a$ to correspond to the moment of time.  The calculations are done up to the moment when the scaling parameter reaches the value $a=6$. As before, \fig{stone-a6} and \fig{stone-v-a6} show the results of simulations at the moment when the scaling parameter becomes equal to $a=6$.

\subsubsection{The 3D rotor problem}
As with the 2D code, we test the 3D code with the rotor problem. The 3D MHD equations are solved numerically in the cylindrical region $r<0.5$, $-0.5<z<0.5$ and the initial conditions are determined using the same initial conditions as in the 2D test. In the beginning of the simulations, $t=0$, the pressure in the region is constant, $p=1$, and the magnetic field is homogeneous, $B_x=0$, $B_y=5$. In the center there is a circle with radius $r_0=0.1$ where the matter density is $\rho_0=10$ and the matter rotates as a solid body with angular velocity $\omega_0=20$. At $r>r_1=0.115$, the density is $\rho_1=1$ and the matter is at rest. In the ring $r_0<r<r_1$, the density and velocity are linearly interpolated between those at $r=r_0$ and $r=r_1$.

\begin{figure}
\includegraphics[width=7cm]{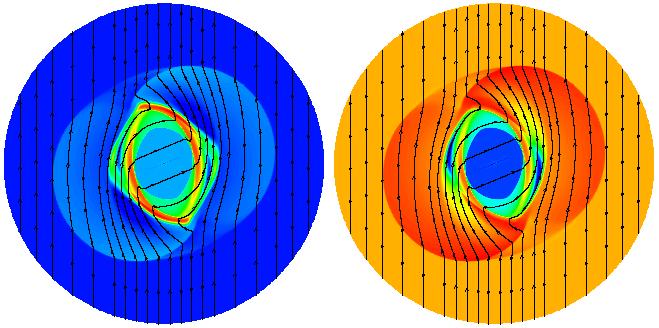}
\caption{Test of the ideal MHD module of the 3D version of code with the ``rotor problem" on a grid with $N_r, N_\phi, N_z = 500 \times 500 \times 20$.  {\it Left Panel:} Density distribution (color background) and field lines (solid lines). {\it Right Panel:} The pressure distribution (color background) and field lines (solid lines).}
\label{rotor}
\end{figure}

In the test we use a homogeneous grid with $N_r, N_\phi, N_z = 500 \times 500 \times 20$. The integration time-step is chosen automatically in such way that $\Delta t = 0.4 \Delta t_{CFL}$. The results of the simulations at time $t=0.12$ are shown in \fig{rotor}. One can see that the density, pressure and the field line distributions are very similar to 2D case.

\subsubsection{3D Low-type analytical solution} 
\begin{figure}
\includegraphics[width=7.cm]{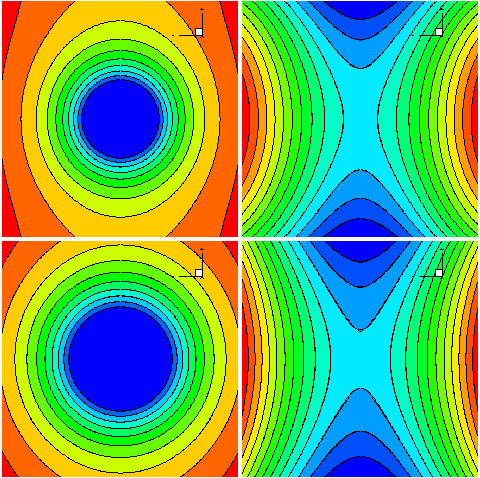}
\caption{Test of the ideal MHD module of the 3D version of code with the Low-type solution test at different times. {\it Top Panels:} Pressure (left panel) and density (right panel) distributions (color background) and analytical Low-type solution (solid lines) with grid resolution $N_r, N_\phi, N_z = 100 \times 6 \times 200$ at scaling parameter $a=2.15$, {\it Bottom Panels:} scaling parameter $a=2.4$.} \label{lou}
\end{figure}

To test the ideal MHD module, we also use the analytical solution described in Sec. \ref{subsubsec:2Dlow}. Calculations were performed in the cylindrical region $r<6,~-6<z<6$ for a grid $N_r, N_\phi, N_z = 100 \times 6 \times 200$. Results of simulations are shown in \fig{lou}. The color background shows pressure distribution (left panel) and density (right panel). The lines show corresponding analytical solutions.

\begin{figure}
\centering
\includegraphics[width=7cm]{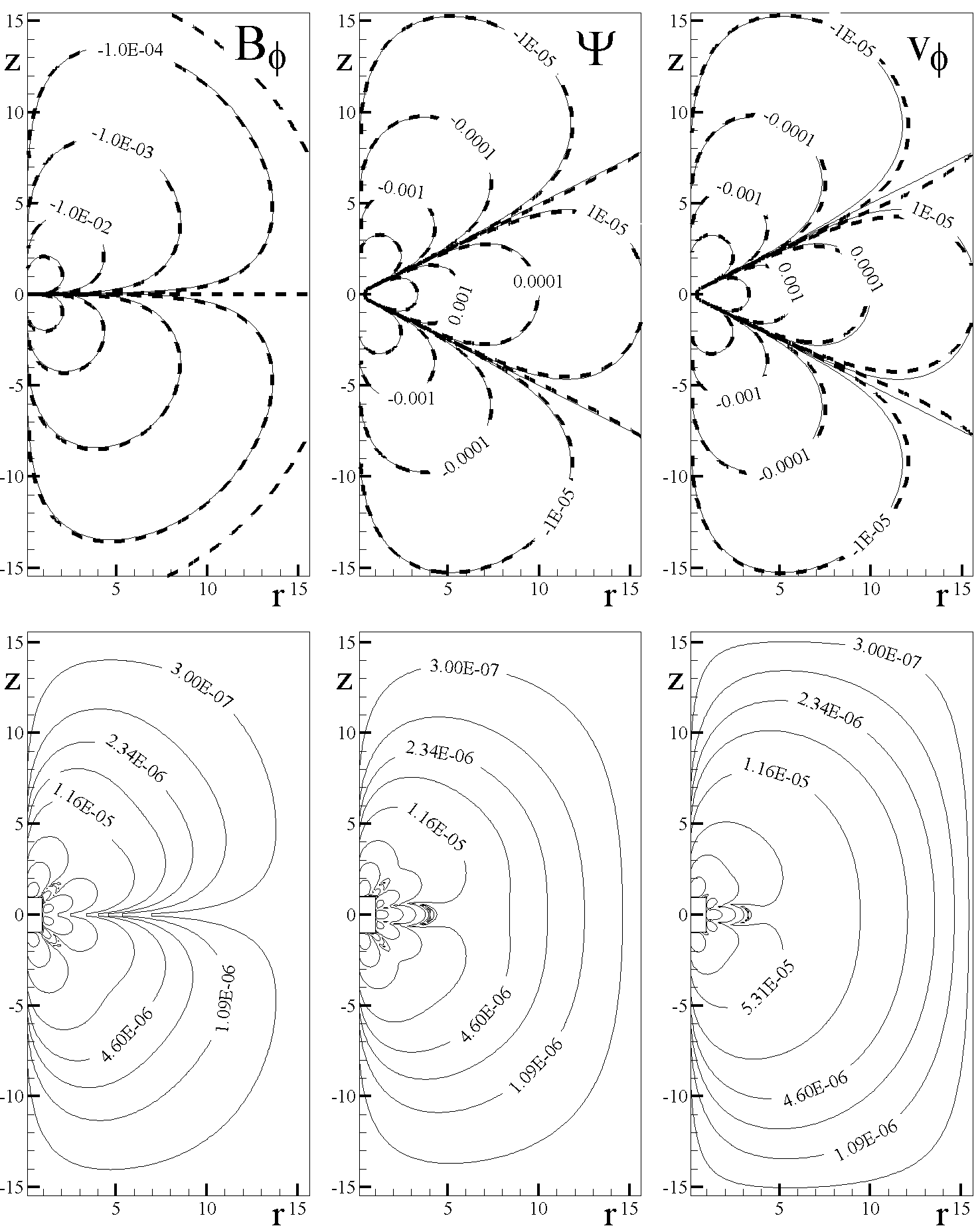}
\caption{The top panels show contours of the azimuthal component of the magnetic field $B_\phi$, magnetic flux function $\Psi$ and the azimuthal velocity $v_\phi$, from left to right. The solid lines indicate the numerical solutions while the dashed lines show the analytical solutions given by Eqs. \ref{5-5}--\ref{5-7}. The bottom panels show the absolute errors in the calculation of the corresponding functions.} \label{diff-50-100}
\end{figure}

\begin{figure*}
\centering
\includegraphics[width=4.in]{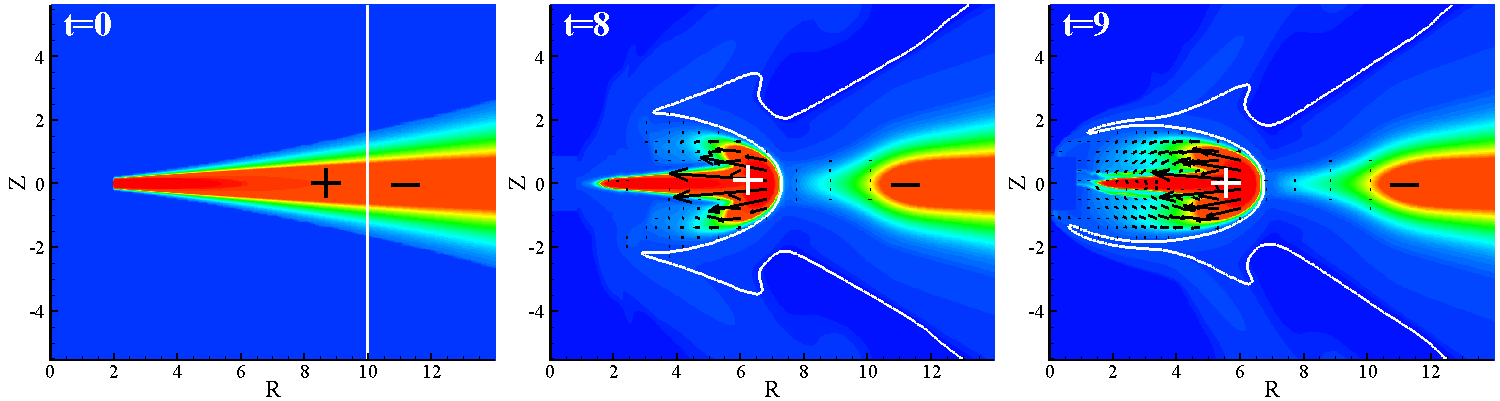}
\includegraphics[width=4.in]{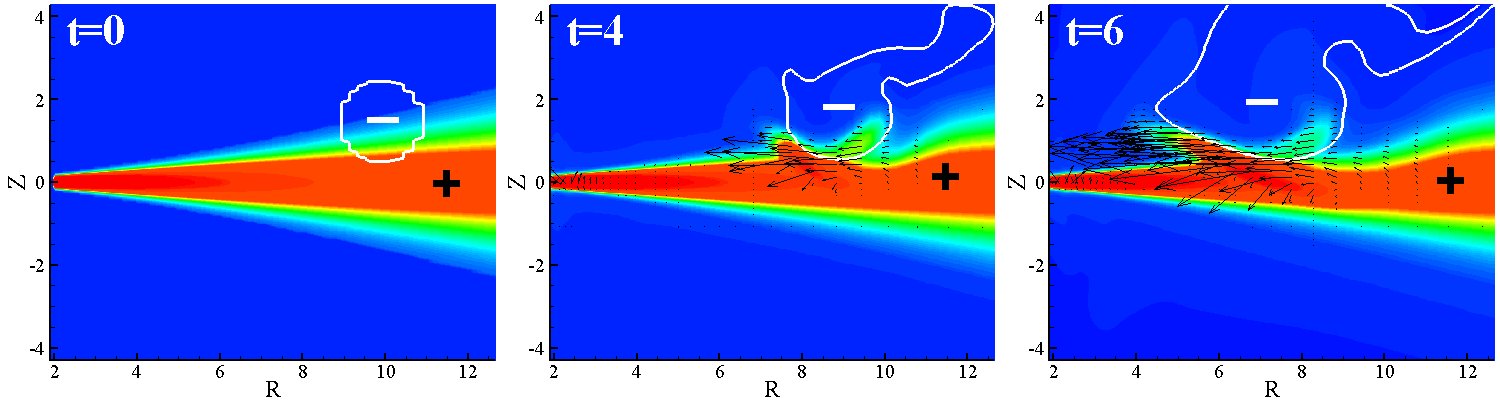}
\caption{An example of hydro simulations of the interaction of the counter-rotating disks. \textit{Top Panels} show the case where the inner and outer parts of the disk are rotating in the opposite directions.  \textit{Bottom Panels} show the case where a clump of gas rotates in the direction opposite to the rotation of the disk.  The color background shows the density distribution, vectors show matter flux, and the white bold line separates the regions of opposite angular velocities which are marked with the ``+" and ``-" signs. Simulations are performed in 2d cylindrical coordinates with the grid $N_r\times N_z=150\times250$. From \cite{DydaEtAl2015}.} \label{counter-blob}
\end{figure*}

\subsection{Test of the diffusivity module }
Here we test the diffusivity module of the code. For this we switch off the hydrodynamic fluxes and the corresponding right-hand side terms in Eqs. \ref{4-4} and \ref{4-5} and integrate these equations numerically in the region $0<r<r_1$, $-r_1 < z < r_1$ with the rectangular region $0<r<r_0$, $-r_0 < z < r_0$ excised from simulations.
\begin{align}
\label{5-2} &\parti{\Psi}{t} - \eta \left( r \parti{}{r}
\frac{1}{r} \parti{\Psi}{r}
+ \partii{\Psi}{z} \right) = 0, \\
\label{5-3} &\parti{B_\phi}{t} -
\parti{}{r} \left( \frac{\eta}{r} \parti{r B_\phi}{r} \right)
+ \parti{}{z} \left( \eta \parti{B_\phi}{z} \right) = 0, \\
\label{5-4} &\parti{(\rho r \omega)}{t} - \frac{1}{r^2}
\parti{}{r} \left( \mu r^3 \parti{\omega}{r} \right) +
\parti{}{z} \left( \mu r \parti{\omega}{z} \right) = 0 ~.
\end{align}
The solutions for the magnetic flux function $\Psi$ and azimuthal velocity $v_\phi$ are symmetric about the equatorial plane while solutions for the azimuthal component of the magnetic field $B_\phi$ are antisymmetric. All solutions go to zero at the symmetry axis
$$
\Psi |_{r=0} = 0,~~
B_\phi |_{r=0} = 0,~~
v_\phi |_{r=0} = 0.
$$
We can find particular solutions of Eqs. \ref{5-2}--\ref{5-4} by the method of variable separation in spherical coordinates. For $\eta = 1$, Eq. \ref{5-2} in spherical coordinates is
\begin{equation}
\label{5-2sph}
\parti{\Psi}{t} = \partii{\Psi}{R} + \sin \theta \parti{}{\theta}
\left( \frac{1}{\sin \theta} \parti{\Psi}{\theta} \right) ~.
\end{equation}
For testing we choose the following solutions of Eq. \ref{5-2sph}:
\begin{align}
\label{5-5} &\Psi=\frac{C \sin^2\theta}{R}\bigg( 1 - \frac{5
\sin^2\theta}{4} \bigg) \bigg( 1+ \frac{10(t_0-t)}{R^2} \bigg), \\
\label{5-6} &B_\phi=\frac{C \sin\theta \cos\theta}{R} \bigg( 1 +
\frac{6(t_0-t)}{R^2} \bigg), \\ \label{5-7} &v_\phi=\frac{ C
\sin\theta}{R^2}\bigg( 1-\frac{5 \sin^2\theta}{4}\bigg )
\bigg(1+\frac{10(t_0-t)}{R^2} \bigg),
\end{align}
where $C$ and $t_0$ are constants, $R = \sqrt{r^2+z^2}$ is the spherical radius, and $\theta$ is the polar angle.

The equations for the magnetic flux function and the $\phi$-components of the velocity and magnetic field are integrated numerically using our 2.5D code (with the switched-off hydrodynamic fluxes and zero right hand side terms).  We use the same grid geometry as in the main simulations; that is, cylindrical coordinates with a non-equidistant grid in both directions.

For the test, we take a grid of $N_r, N_z = 50 \times 100$, and also a finer grid with $N_r, N_z = 100 \times 200$. We also take the inner and outer radii (in the $r-$direction) of the simulation to be $r_0=1$ and $r_1=15.6$. For these boundaries we set $\Psi, B_\phi$ and $v_\phi$ to be equal to the analytical values determined by Eqs. \ref{5-5}-\ref{5-7}. The initial conditions correspond to the exact solutions of Eqs. \ref{5-5}-\ref{5-7} at $t=0$. The constants $C$ and $t_0$ in these equations are chosen so that the solutions at the inner boundary are of the order of unity. We use $C=0.01$, $t_0=50$ and integrate the equations up to $t_1 = 2 t_0 = 100$.

\begin{table}
\centering
\begin{tabular}{l@{\extracolsep{0.2em}}l@{}lll}
\hline
&   Grids                     & $\max|\Delta \Psi|$     & $\max|\Delta B_\phi|$  & $\max|\Delta v_\phi|$  \\
\hline \multicolumn{2}{l}{$50\times 100$}     & $3.68\e{-2}$     & $3.19\e{-2}$
& $7.59\e{-2}$       \\
\multicolumn{2}{l}{$100\times 200$}    & $1.01\e{-2}$      & $6.51\e{-3}$
& $1.60\e{-2}$       \\
\hline
\end{tabular}
\caption{The maximum absolute error in calculations of the $\Psi, B_\phi $ and $v_\phi$ obtained in the diffusivity
tests.} \label{t:dif}
\end{table}

The top panels of \fig{diff-50-100} show the result of the diffusivity test for the $\phi$-component of the magnetic field, the vector potential, and the $\phi$-component of the velocity.  The solid lines show the numerical solution of Eqs. \ref{5-2}-\ref{5-4}, which are obtained by integrating the equations using our code until $t=100$. The dashed lines show the exact solutions given by Eqs. \ref{5-5}-\ref{5-7}. Evidently, the numerical and analytical solutions are nearly identical. The bottom panel shows the absolute errors $\Delta \Psi={\Psi_{\rm num} - \Psi_{\rm exact}}$ for each function. One can see that the error is very small everywhere, taking into account the fact that the maximum value of all the functions, $\Psi,~B_\phi,~v_\phi \approx 1$. The error increases towards the star because the value of the functions also strongly increases towards the star.  We perform similar simulations and analysis for finer grid resolutions $100\times 200$. Table \ref{t:dif} shows the maximum absolute errors, $\max|\Psi_{\rm num} - \Psi_{\rm exact}|$, $\max|B_{\phi}^{num} - B_{\phi}^{exact}|$, and $\max|v_{\phi}^{num} - v_{\phi}^{exact}|$. One can see that for the finer grid (dimensions $100 \times 200$), this error is $2.1$ and $3.9$ times smaller compared to the coarsest grid (dimensions $50 \times 100$). This confirms the convergence of the numerical solution towards the analytical solution. Similar comparisons for the $B_\phi$ component give factors $2.2$ and $4.1$, which also show convergence. Similar comparisons for the $v_\phi$ component give factors $2.2$ and $4.1$, which also show convergence.

\section{Examples of Astrophysical Simulations}
\label{sec:astrophysical examples}

The codes discussed in this paper have already been used to study a variety of different astrophysical problems. Below, we show a few examples of calculations with the code in both 2D and 3D geometries.

\subsection{Counter-rotating accretion disks: 2.5D hydro simulations in cylindrical coordinates}
The hydrodynamic version of the code in 2.5D cylindrical coordinates was used to study counter-rotating accretion disks \citep{DydaEtAl2015}. The simulations show that the interaction between two counter-rotating regions of the disk strongly depends on the value of the viscosity coefficient $\alpha$ which varies from $\alpha=0.02$ up to $\alpha=0.4$. The simulations show that in counter-rotating disks, the accretion rate may be increased by a factor of $\sim10^2-10^4$.

The problem is set up such that the inner region of the disk orbits the star in one direction while the outer part orbits in the opposing direction (see Fig. \ref{counter-blob}, top panels).  The simulations were performed for a viscosity parameter of $\alpha=0.1$. Initially, the boundary between the counter-rotating regions of the disk is sharp. When the simulations start, the oppositely directed regions of the disk start interacting due to viscosity, leading to annihilation of the angular momentum at the interface. This matter, which has lost some of its centrifugal support, begins to accrete rapidly due to the unbalanced gravitational force. This leads to a hundred-fold increase of the accretion rate compared with a disk which does not have a counter-rotating component. The simulations also show that the $r\phi$ component of the stress tensor, $\tau_{r\phi}$, dominates the angular momentum transport in the disk.

We also study the case where a clump of counter-rotating matter is allowed to interact with the top of the disk (see Fig.  \ref{counter-blob}, bottom panels). We find that a part of the disk matter loses angular momentum through viscous interactions with the counter-rotating clump and is quickly accreted onto the star. Analysis shows that in this case, three components of the viscosity tensor $\tau_{r\phi}$, $\tau_{z\phi}$ and $\tau_{rz}$ contribute to the angular momentum transport.

\subsection{MRI-driven accretion onto a magnetized star: 2.5D MHD simulations in cylindrical coordinates}
\label{sec:mri accretion}

\begin{figure*}
\centering
\includegraphics[width=4.5in]{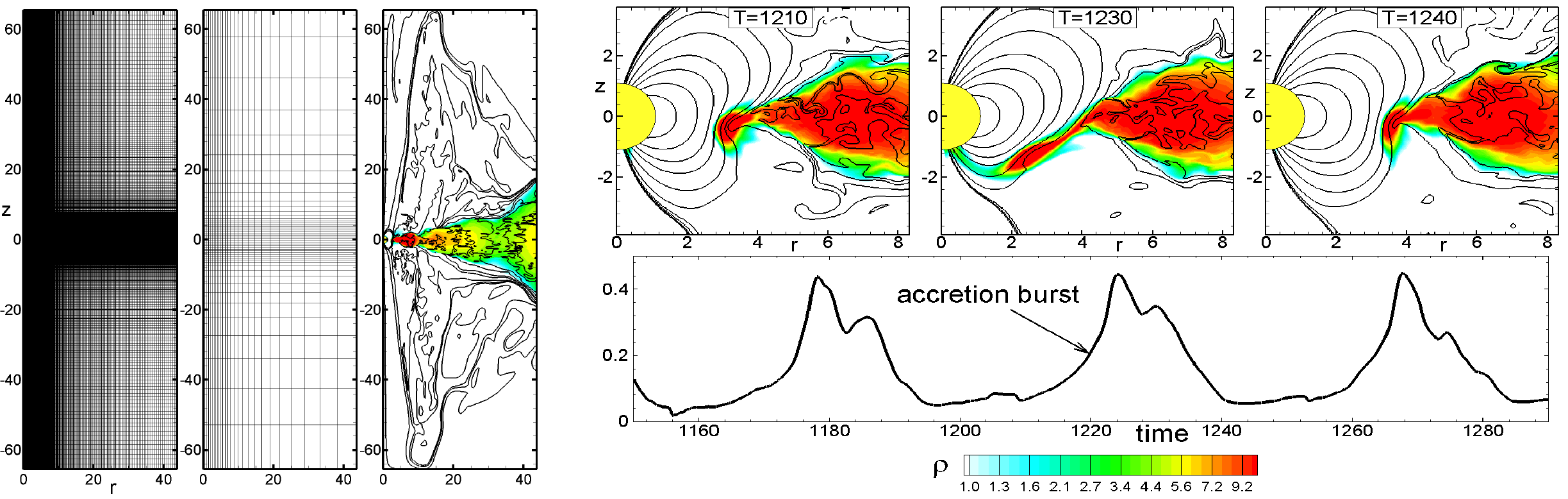}
\caption{From left to right: (1) The cylindrical grid $N_r\times N_z=270\times 432$ cells used in simulations of the magnetospheric accretion, compressed near the axis and in the disk; (2) The same grid, rarefied by a factor of 10, for demonstrative purposes; (3) An example of a simulation of the disk-magnetosphere interaction, where color shows the density distribution, and black lines are magnetic field lines; (4-6, top panels) An example of the disk-magnetosphere interaction in the case of an MRI-driven turbulent disk, the bottom panel shows the accretion rate.  From \citet{RomanovaEtAl2011}.} \label{2d-mri-2}
\end{figure*}

The code has also been utilized to study accretion onto a magnetized star from a turbulent accretion disk in 2.5D cylindrical coordinates \citep{RomanovaEtAl2011}. A sample of the results are shown in the right panel of Figure \ref{2d-mri-2}. The turbulence present in the disk is driven by the magneto-rotational (MRI) instability (e.g., \citealt{BalbusHawley1991}). Simulations were performed for different magnetospheric radii $r_m$ and different initial values of the plasma parameter $\beta=8\pi p/B^2$ in the disk, which varied from $\beta=10$ to $\beta=10,000$. The dynamical range of initial densities between the disk and corona was $10^4$. Inside the magnetosphere, the dipole magnetic field varies as $\sim r^{-3}$ and the magnetic pressure varies as $\sim r^{-6}$. In the regions of high magnetic pressure, the matter pressure and density become very low, thereby limiting the simulation timestep and slowing the calculation. To overcome this difficulty, we require: $p > \beta_{\rm floor} \times B^2/8\pi$, where $\beta_{\rm floor}=10^{-3}$; in other words, we require that the matter pressure does not fall below a certain fraction of the magnetic pressure. In the grid cells where the matter pressure becomes smaller than this value, it is adjusted to be $p=\beta_{\rm floor} \times B^2/8\pi$. We then recalculate the density $\rho=(p/s)^{1/\gamma}$ to be thermodynamically consistent. In cases of large magnetospheres, this leads to slightly enhanced matter density and pressure near the star. However, this does not change the physics of the magnetosphere because the magnetic pressure is still $10^3$ times higher than the matter pressure, and therefore the physics of the disk-magnetosphere interaction is described correctly in the sense that the magnetosphere is strongly magnetically-dominated. In other parts of simulation region, no floor density or pressure is added.

The grid is compressed towards the disk midplane and towards the $z-$axis with the goal of having higher grid resolution in the disk and near the star. The left panel of Figure \ref{2d-mri-2} shows a typical grid used in this study with $N_r=270$ cells in the radial direction and $N_z=432$ cells in the axial direction. The number of grid cells covering the disk in the vertical direction (in the middle of the disk) is about 200, and the MRI-unstable modes are well-resolved even in cases of a small-scale turbulence.

For this study, the viscosity or diffusivity terms are not enabled. The disk matter is loaded onto the magnetic field lines of the magnetosphere due to reconnection or due to the small numerical diffusivity which is present in the code.

\subsection{Propeller outflows from the disk-magnetosphere
boundary: 2.5D MHD simulations}

\begin{figure*}
\centering
\includegraphics[width=4.5in]{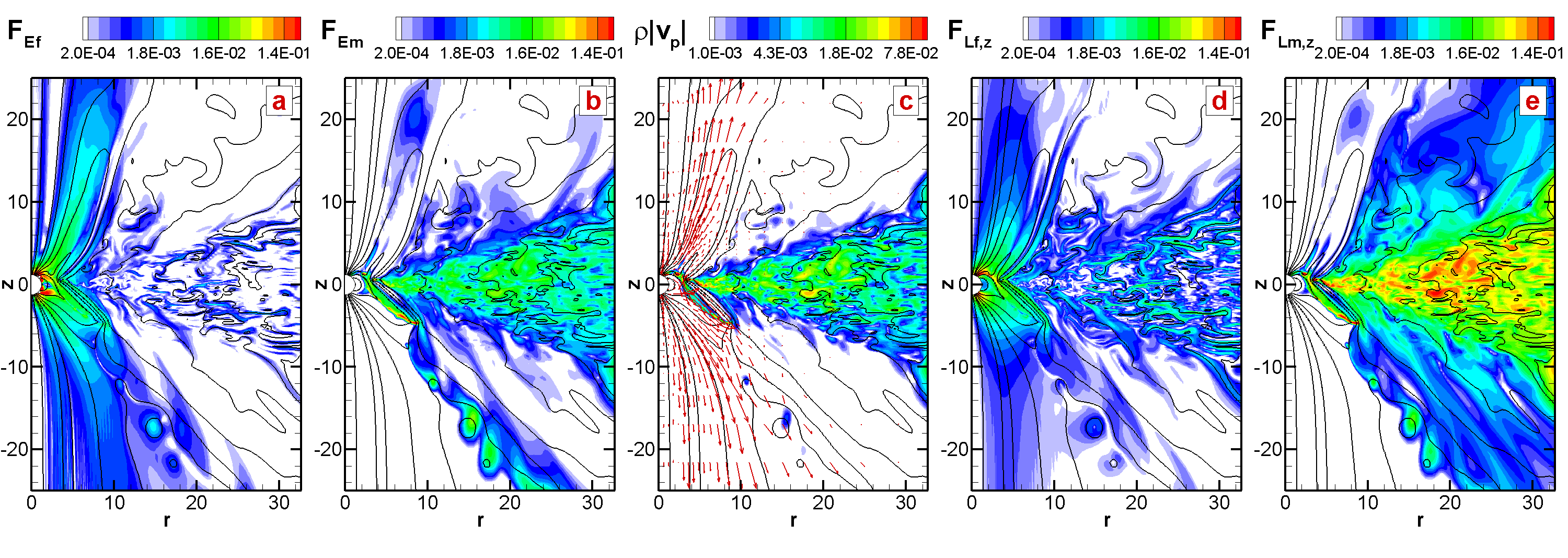}
\caption{An example of a 2D MHD simulation of propeller driven outflows performed in cylindrical coordinates with non-uniform grid of dimension $N_r\times N_z=250 \times 432$ cells.  {\it Panels {\bf a} and {\bf b}:} energy flux density carried by the magnetic fields, ${\bf F}_{Ef}$, and by matter, ${\bf F}_{Em}$.  The lines show the magnetic field lines.  {\it Panel {\bf c}:} the contours show the poloidal matter flux, ${\rho}{v}_{\rm p}$, and the red arrows show the poloidal velocity vectors.  {\it Panels {\bf d} and {\bf e}:} angular momentum flux density carried by the magnetic fields, ${\bf F}_{Lf,z}$, and matter, ${\bf F}_{Lm,z}$.
   From \cite{LiiEtAl2014}.} \label{2d-propeller}
\end{figure*}

Another problem which has been studied using this code is that of disk accretion onto a magnetized star rotating in the propeller regime \citep{LiiEtAl2014}. In the propeller regime of accretion, the magnetosphere of the star rotates more rapidly than the inner disk; the angular momentum imparted on the disk by the magnetosphere can push the disk away or drive outflows from the disk-magnetosphere boundary (e.g., \citealt{IllarionovSunyaev1975, LovelaceEtAl1999}). As in the previous study (Sec. \ref{sec:mri accretion}), the angular momentum transport in the disk is mediated by the MRI-driven turbulence.

The large centrifugal barrier at the disk-magnetosphere boundary inhibits direct accretion onto the star. Rather, the disk matter accumulates at the disk-magnetosphere boundary until it compresses the magnetosphere inward of the corotation radius and accretion can proceed. The result is a burst of accretion in which part of the matter accretes onto the star and part is launched as a short-lived magneto-centrifugal wind. There is also a secondary low-density wind which flows out along field lines anchored to the surface of the star. In the latter component of the winds, the matter is launched rapidly, resulting in low density regions near the star.

In order to prevent the density (and simulation timestep) from becoming too small, we implement a floor density condition. That is, we fix the minimum density in the simulation region to be $\rho_{floor}=0.01 \rho_{corona}=10^{-6}$, effectively adding a small amount of matter whenever the density drops below a certain threshold. We track the amount of matter that is added, and find that it is much smaller than the matter flux in the disk and in the wind. We also implement the matter pressure floor condition for matter pressure inside the magnetosphere (see Sec.  \ref{sec:mri accretion}).

As a word of caution: for relatively large values of $\beta_{\rm floor}$, the matter which is added to maintain the floor pressure can have a tendency to flow out of the super-Keplerian magnetosphere due to centrifugal forces, thereby forming an artificial outflow. However, for low enough values of $\beta_{\rm floor}$ (up to $\beta_{\rm floor}=3\times 10^{-5}$), no artificial outflows from the inner magnetosphere were observed.

Fig. \ref{2d-propeller} shows matter flux and also energy and angular momentum carried by the matter and the magnetic field (see details and description of fluxes in \citealt{LiiEtAl2014}). The axisymmetric grid is in cylindrical ($r$, $z$) coordinates with mesh compression towards the disk and towards the $z$-axis such that there are a larger number of cells in the disk plane and near the star. In the models presented here, we use a non-uniform grid with dimension $250 \times 432$ cells.  corresponding to a grid that is 66 by 140 stellar radii in size.  At $r=20$, the number of grid cells which cover the disk in the vertical direction is about 200.

At the disk-magnetosphere boundary, 3D instabilities may cause the disk matter to penetrate through the magnetic field lines into the magnetosphere. Since these instabilites are intrinsically three dimensional, they are not modeled in the current axisymmetric 2.5D simulations. To mimic these processes, we implement diffusivity near the star using the diffusivity module described in Sec.  \ref{subsec:viscosity and diffusivity}. To prevent modifications to the overall disk accretion rate, the diffusivity is only enabled near the disk-magnetosphere boundary. We study the disk-magnetosphere interaction in this region for different diffusivity parameters $\alpha_d=0.01, 0.03, 0.1$. We find that for larger values of $\alpha_d$, the disk matter penetrates through the magnetosphere more rapidly causing most of the accreting matter to be ejected into an outflow.

\subsection{Migration of a planet in the magnetized disk: 2D MHD simulations in polar coordinates}

\begin{figure*}
\centering
\includegraphics[width=3.60in]{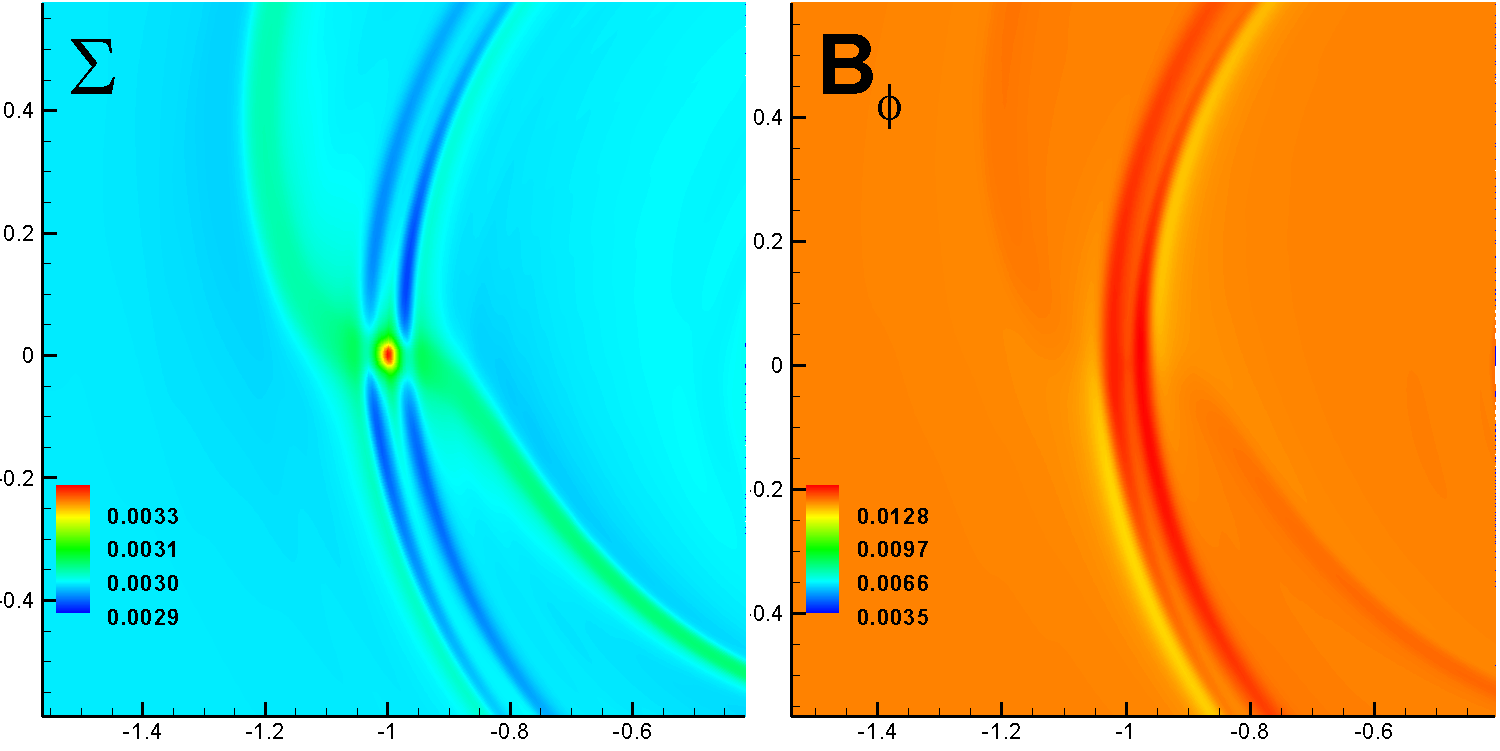}
\includegraphics[width=1.4in]{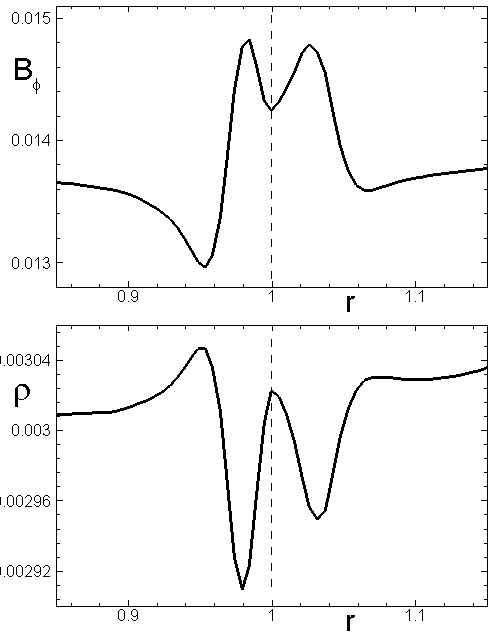}
\caption{Modeling a planet in a magnetized disk in 2d polar coordinates. The planet excites slow magnetosonic waves at the magnetic resonances which are seen as low-density rings in the left panel and an enhanced azimuthal field in the middle panel.  The right panel shows the linear distribution of the surface density and magnetic field in the region of the resonances.  Simulations performed with the grid $N_r=480$, and $N_\phi=1200$.  From Comins et al. (in prep.).} \label{magres}
\end{figure*}

A version of the code in 2D polar coordinates has been used to model the migration of a planet through a magnetized disk (Comins et al. 2015, in prep.).  In a test version of this problem, we followed the approach by \citet{FromangEtAl2005} which used two different codes to solve this problem (\textit{FARGO} and another code). Our initial conditions are almost identical to those taken by \citet{FromangEtAl2005}: the planet's mass is 5 Earth masses; the disk has a constant surface density distribution; the initial magnetic field has only an azimuthal component, $B_\phi$, which scales with radius as $B_\phi=B_{\phi0} (r_0/r)^k$ where $k=0, 1, 2$; and lastly, the disk is locally-isothermal. We take $\gamma=1.01$ for the adiabatic index to mimic a locally-isothermal disk. The size of simulation region is $R_{min}=0.4$ and $R_{max}=4.9$, and the grid $N_r=480$ and $N_\phi=1200$.  The grid is stretched in the radial direction such that the grid cells are almost rectangular in all parts of the simulation region. The magnetic resonances are resolved at 12 grid cells in the radial direction.

Our simulations confirm the results obtained by \citet{FromangEtAl2005}. Namely, that if the plasma parameter $\beta=8\pi p/B^2$ is of order unity, then the magnetic resonances can be more important than the Lindblad resonances (which lead to inward migration) and the migration can be stopped if the radial gradient of the azimuthal magnetic field is sufficiently large. Fig. \ref{magres} shows the density distribution where the magnetic resonances are observed as azimuthally-stretched density depletions near the planet (see left panel). The middle panel shows that in the region of magnetic resonances, the magnetic field is enhanced. The right panel shows a radial cut of the density and magnetic field along the planet's radial vector. We find that at $k=0$ the migration is inward; in contrast, for steeper density profiles ($k=1$ and $k=2$) the planet migrates outward. This is in accord with the results obtained by \citet{FromangEtAl2005}, who find that the inward migration consistently slows and reverses direction as $k$ increases.

\subsection{Migration of a planet in non-magnetized laminar disk: 3D hydro simulations in cylindrical coordinates}

\begin{figure*}
\centering
\includegraphics[width=4.0in]{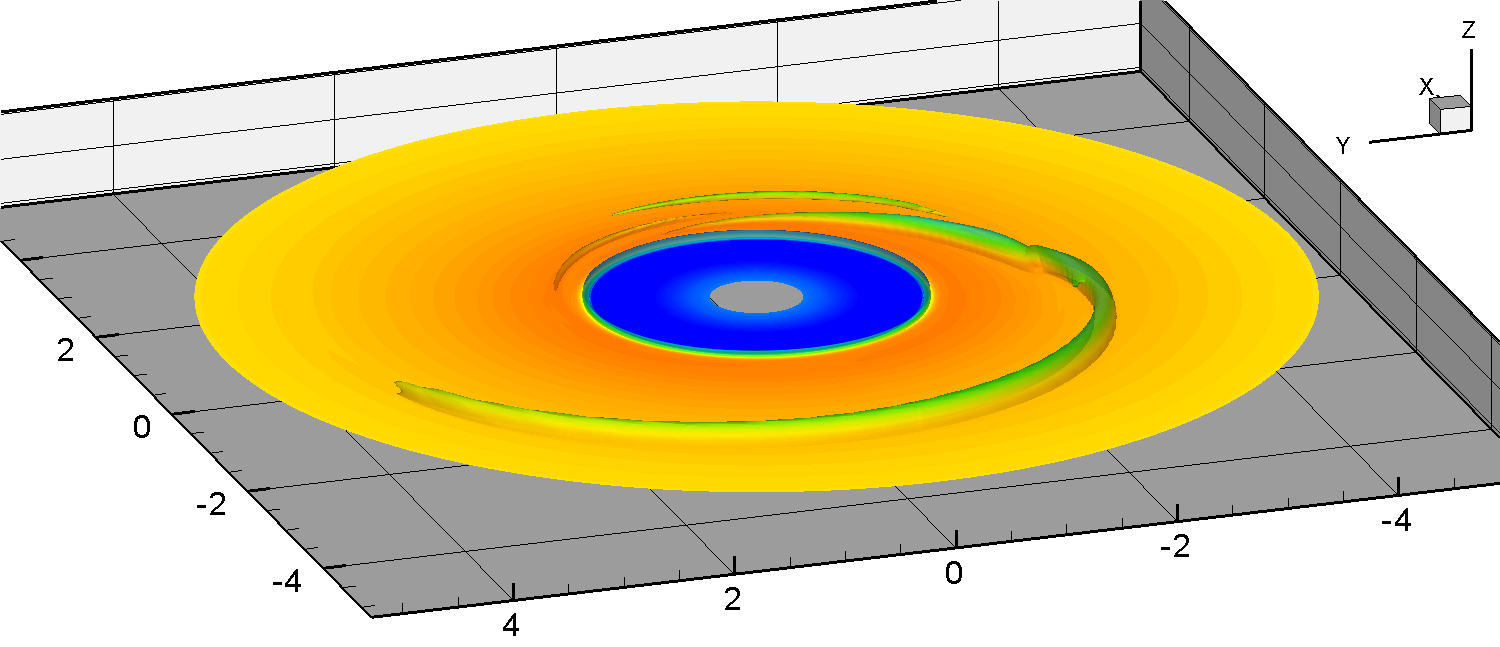}
\caption{An example of a 3D simulation of a 15 Earth mass planet migrating in a thin disk. The cut shows the equatorial density
distribution and the translucent 3D density iso-level shows the Lindblad waves. Simulations performed with the grid $N_r=232$,
$N_\phi=480$ and $N_z=80$.}
\label{3d-planet}
\end{figure*}

A version of the 3D code in cylindrical coordinates has been used to model the migration of a planet in a 3D hydrodynamic disk. The cylindrical simulation region extends between $R_{min}=0.4$ and $R_{max}=4.9$ and $Z_{max}=\pm 0.5$. The number of grid cells in each direction is $N_r=232$, $N_\phi=480$, and $N_z=80$. The simulations show that the planet excites two waves in the disk at the Lindblad resonances and migrates inward due to angular momentum transfer from the planet to the disk. The density perturbation is largest in the equatorial plane, and falls off away from the equatorial plane. Fig. \ref{3d-planet} shows an example of simulations which show typical waves in the disk.

\section{Conclusions}
\label{sec:conclusions}

We have developed and tested a Godunov-type code where the HLLD solver of Miyoshi and Kusano \citep{MiyoshiKusano2005} is implemented.  Due to the nature of many of our astrophysical problems (where strong shock waves are not expected), we choose the adiabatic form of the energy equations. The codes pass standard tests and were used for solving different astrophysical problems.

The simulations show that our entropy-conserving HLLD solver performs satisfactorily in many astrophysical applications, such as studies of low-density, high-velocity outflows in propeller-driven winds. The ideal MHD module, viscosity module and diffusivity module were tested in different astrophysical situations.

\section*{Acknowledgments} 
The authors thank the anonymous referee for valuable comments and constructive criticism and Alisa Blinova for help with this paper. Resources supporting this work were provided by the NASA High-End Computing (HEC) Program through the NASA Advanced Supercomputing (NAS) Division at the NASA Ames Research Center and the NASA Center for Computational Sciences (NCCS) at Goddard Space Flight Center. The research was supported by NASA grant NNX12AI85G and NSF grant AST-1211318. AVK was supported by the Russian academic excellence project ``5top100".

\end{document}